\documentclass[dvips]{article}
\usepackage{supertabular,lscape,epsfig}
\usepackage{amssymb}
\usepackage{amsmath}
\usepackage{rotating}
\DeclareSymbolFont{ppa}{OT1}{ppl}{m}{it}
\DeclareMathSymbol{\vv}{\mathalpha}{ppa}{'166}

\begin{document}

\newcommand{\dd}{\,{\rm d}}
\newcommand{\ie}{{\it i.e.},\,}
\newcommand{\etal}{{\it et al.\ }}
\newcommand{\eg}{{\it e.g.},\,}
\newcommand{\cf}{{\it cf.\ }}
\newcommand{\vs}{{\it vs.\ }}
\newcommand{\zdot}{\makebox[0pt][l]{.}}
\newcommand{\up}[1]{\ifmmode^{\rm #1}\else$^{\rm #1}$\fi}
\newcommand{\dn}[1]{\ifmmode_{\rm #1}\else$_{\rm #1}$\fi}
\newcommand{\upd}{\up{d}}
\newcommand{\uph}{\up{h}}
\newcommand{\upm}{\up{m}}
\newcommand{\ups}{\up{s}}
\newcommand{\arcd}{\ifmmode^{\circ}\else$^{\circ}$\fi}
\newcommand{\arcm}{\ifmmode{'}\else$'$\fi}
\newcommand{\arcs}{\ifmmode{''}\else$''$\fi}
\newcommand{\MS}{{\rm M}\ifmmode_{\odot}\else$_{\odot}$\fi}
\newcommand{\RS}{{\rm R}\ifmmode_{\odot}\else$_{\odot}$\fi}
\newcommand{\LS}{{\rm L}\ifmmode_{\odot}\else$_{\odot}$\fi}
\newcommand{\feh}{\hbox{$ [{\rm Fe}/{\rm H}]$}}

\newcommand{\Abstract}[2]{{\footnotesize\begin{center}ABSTRACT\end{center}
\vspace{1mm}\par#1\par
\noindent
{~}{\it #2}}}

\newcommand{\TabCap}[2]{\begin{center}\parbox[t]{#1}{\begin{center}
  \small {\spaceskip 2pt plus 1pt minus 1pt T a b l e}
  \refstepcounter{table}\thetable \\[2mm]
  \footnotesize #2 \end{center}}\end{center}}

\newcommand{\TableSep}[2]{\begin{table}[p]\vspace{#1}
\TabCap{#2}\end{table}}

\newcommand{\FigCap}[1]{\footnotesize\par\noindent Fig.\  %                     
  \refstepcounter{figure}\thefigure. #1\par}

\newcommand{\TableFont}{\footnotesize}
\newcommand{\TableFontIt}{\ttit}
\newcommand{\SetTableFont}[1]{\renewcommand{\TableFont}{#1}}

\newcommand{\MakeTable}[4]{\begin{table}[htb]\TabCap{#2}{#3}
  \begin{center} \TableFont \begin{tabular}{#1} #4
  \end{tabular}\end{center}\end{table}}

\newcommand{\MakeTableSep}[4]{\begin{table}[p]\TabCap{#2}{#3}
  \begin{center} \TableFont \begin{tabular}{#1} #4
  \end{tabular}\end{center}\end{table}}

\newenvironment{references}%                                                    
{
\footnotesize \frenchspacing
\renewcommand{\thesection}{}
\renewcommand{\in}{{\rm in }}
\renewcommand{\AA}{Astron.\ Astrophys.}
\newcommand{\AAS}{Astron.~Astrophys.~Suppl.~Ser.}
\newcommand{\ApJ}{Astrophys.\ J.}
\newcommand{\ApJS}{Astrophys.\ J.~Suppl.~Ser.}
\newcommand{\ApJL}{Astrophys.\ J.~Letters}
\newcommand{\AJ}{Astron.\ J.}
\newcommand{\IBVS}{IBVS}
\newcommand{\PASJ}{PASJ}
\newcommand{\PASP}{P.A.S.P.}
\newcommand{\Acta}{Acta Astron.}
\newcommand{\MNRAS}{MNRAS}
\renewcommand{\and}{{\rm and }}
\section{{\rm REFERENCES}}
\sloppy \hyphenpenalty10000
\begin{list}{}{\leftmargin1cm\listparindent-1cm
\itemindent\listparindent\parsep0pt\itemsep0pt}}%                               
{\end{list}\vspace{2mm}}

\def\TYLDA{~}
\newlength{\DW}
\settowidth{\DW}{0}
\newcommand{\dw}{\hspace{\DW}}

\newcommand{\refitem}[5]{\item[]{#1} #2%                                        
\def\REFARG{#3}\ifx\REFARG\TYLDA\else, {\it#3}\fi
\def\REFARG{#4}\ifx\REFARG\TYLDA\else, {\bf#4}\fi
\def\REFARG{#5}\ifx\REFARG\TYLDA\else, {#5}\fi.}

\newcommand{\Section}[1]{\section{#1}}
\newcommand{\Subsection}[1]{\subsection{#1}}
\newcommand{\Acknow}[1]{\par\vspace{5mm}{\bf Acknowledgments.} #1}
\pagestyle{myheadings}

\newfont{\bb}{ptmbi8t at 12pt}
\newcommand{\xrule}{\rule{0pt}{2.5ex}}
\newcommand{\xxrule}{\rule[-1.8ex]{0pt}{4.5ex}}
\def\thefootnote{\fnsymbol{footnote}}

\begin{center}
{\Large\bf Large Variety of New Pulsating Stars\\
in the OGLE-III Galactic Disk Fields\footnote{Based on observations
obtained with the 1.3-m Warsaw telescope at the Las Campanas
Observatory of the Carnegie Institution for Science.}}
\vskip1cm
{\bf
P.~~P~i~e~t~r~u~k~o~w~i~c~z$^1$,~~W.~A.~~D~z~i~e~m~b~o~w~s~k~i$^1$,~~P.~~M~r~\'o~z$^1$,\\
~~I.~~S~o~s~z~y~\'n~s~k~i$^1$,~~A.~~U~d~a~l~s~k~i$^1$,~~R.~~P~o~l~e~s~k~i$^{1,2}$,\\
~~M.~K.~~S~z~y~m~a~\'n~s~k~i$^1$,~~M.~~K~u~b~i~a~k$^1$,~~G.~~P~i~e~t~r~z~y~\'n~s~k~i$^{1,3}$,\\
~~{\L}.~~W~y~r~z~y~k~o~w~s~k~i$^{1,4}$,~~K.~~U~l~a~c~z~y~k$^1$,\\
~~S.~~K~o~z~{\l}~o~w~s~k~i$^1$~~and~~J.~~S~k~o~w~r~o~n$^1$\\}
\vskip3mm
{
$^1$Warsaw University Observatory, Al. Ujazdowskie 4, 00-478 Warszawa, Poland\\
e-mail:
(pietruk,wd,pmroz,soszynsk,udalski,rpoleski,msz,mk,pietrzyn,\\
wyrzykow,kulaczyk,simkoz,jskowron)@astrouw.edu.pl\\
$^2$ Department of Astronomy, Ohio State University, 140 W. 18th Ave.,\\
Columbus, OH 43210, USA\\
$^3$ Universidad de Concepci{\'o}n, Departamento de Astronom\'ia,\\
Casilla 160-C, Concepci{\'o}n, Chile\\
$^4$ Institute of Astronomy, University of Cambridge, Madingley Road,\\
Cambridge CB3 0HA, UK\\
}
\end{center}

\Abstract{We present the results of a search for pulsating stars in the
7.12 deg$^2$ OGLE-III Galactic disk area in the direction tangent to the
Centaurus Arm. We report the identification of 20 Classical Cepheids,
45 RR~Lyr type stars, 31 Long-Period Variables, such as Miras
and Semi-Regular Variables, one pulsating white dwarf, and 58 very likely
$\delta$~Sct type stars. Based on asteroseismic models constructed for one
quadruple-mode and six triple-mode $\delta$~Sct type pulsators, we estimated
masses, metallicities, ages, and distance moduli to these objects.
The modeled stars have masses in the range 0.9--2.5~$M_{\odot}$ and are
located at distances between 2.5~kpc and 6.2~kpc. Two triple-mode
and one double-mode pulsators seem to be Population II stars
of the SX~Phe type, probably from the Galactic halo. Our sample
also includes candidates for Type II Cepheids and unclassified
short-period ($P<0.23$~d) multi-mode stars which could be either
$\delta$~Sct or $\beta$~Cep type stars. One of the detected variables
is a very likely $\delta$~Sct star with an exceptionally high peak-to-peak
$I$-band amplitude of 0.35~mag at the very short period of 0.0196~d.
All reported pulsating variables but one object are new discoveries.
They are included in the OGLE-III Catalog of Variable Stars.

Finally, we introduce the on-going OGLE-IV Galactic Disk Survey,
which covers more than half of the Galactic plane. For the purposes of
future works on the spiral structure and star formation history
of the Milky Way, we have already compiled a list of known Galactic
Classical Cepheids.}

{Galaxy: disk -- Stars: variables: Cepheids -- Stars: variables: delta
Scuti -- Stars: variables: RR~Lyrae -- variables: general}

%%%%%%%%%%%%%%%%%%%%%%%%%%%%%%%%%%%%%%%%%%%%%%%%%%%%%%%%%%%%%%%%%

\Section{Introduction}

Pulsations are present in many phases of stellar evolution and are
reflected in a variety of observed types of pulsating variable stars.
Pulsating stars occupy different regions on the Hertzsprung--Russell
diagram. They are seen among main-sequence (MS) and post-MS stars
(\eg $\beta$~Cep, $\gamma$~Dor, $\delta$~Sct type), giants (\eg Classical
Cepheids, Miras, RR~Lyr type), and white dwarfs (\eg ZZ~Cet, V777 Her type).

Pulsations provide information on the structure and evolution of stars.
The observed properties of Classical Cepheids ($\delta$~Cep type stars),
such as periods, amplitudes, and colors allow for testing of hydrodynamical
models of stars (\eg Kanbur, Ngeow and Buchler 2004, Kanbur \etal 2010).
Period changes in Classical Cepheids help to verify their evolutionary models
(\eg Pietrukowicz 2003, Turner, Abdel-Sabour Abdel-Latif and Berdnikov 2006,
Poleski 2008). Recent studies of period changes show that some Cepheids,
including the nearest Polaris, are very likely undergoing enhanced mass loss
(Neilson \etal 2012).

Better understanding of Cepheid structure provides greater insight into
their use as standard candles. They serve as distance indicators
to various regions of the Milky Way and to other galaxies in the Local Group
and beyond. Classical Cepheids, as young and luminous stars, help to explore
the distribution of elements in the Galactic disk (Luck \etal 2011)
and trace its spiral structure (\eg Popova 2006, Majaess, Turner and Lane 2009,
Bobylev and Bajkova 2012). Recent works indicate a steady decrease
in metallicity in Cepheids with the increasing distance from the Galactic
center in the thin disk and no correlation between the metallicity and
the vertical distance from the Galactic plane (Genovali \etal 2013,
Lemasle \etal 2013).

Numerous Classical Cepheids detected in the Large Magellanic Cloud (\eg\\
Soszy\'nski \etal 2008) were used for calibration of the extra-galactic
distance scale (\eg Storm \etal 2011). The Cepheid period-luminosity
relation allows for the determination of the distance to many nearby
galaxies, such as M31 (\eg Vilardell, Jordi and Ribas 2007,
Riess, Fliri and and Valls-Gabaud 2012), M33 (\eg Gieren \etal 2013),
NGC 300 (\eg Gieren \etal 2004), IC 1613 (\eg Tammann, Reindl and Sandage 2011),
and more distant, like M81 (Freedman \etal 1994).

RR~Lyr type stars are tracers of old populations. So far, thousands
of such stars have been discovered in the Galactic bulge
(\eg Soszy\'nski \etal 2011a), halo (\eg Szczygie{\l}, Pojma\'nski
and Pilecki 2009, Sesar \etal 2010, Akhter \etal 2012, S\"uveges \etal 2012),
and thick disk (\eg Kinemuchi \etal 2006, Bernhard and Wils 2009,
Mateu \etal 2012). Observations of Galactic halo
variables show the presence of two groups having different metal content
and indicating that the halo was formed by at least two distinct accretion
processes (Miceli \etal 2008). One of unsolved issues is whether the bulge
and metal-rich halo RR~Lyr stars constitute the same population
(Pietrukowicz \etal 2012). RR~Lyr variables serve as distance indicators
and probes of formation history in nearby galaxies (\eg Pietrzy\'nski
\etal 2008, Greco \etal 2009, Sarajedini \etal 2009, Yang \etal 2010).

Another class of pulsating giants, Mira stars, are also used
for measuring distances within the local universe (\eg Whitelock \etal 2009,
Menzies \etal 2010). These bright high-amplitude Long-Period Variable
(LPV) stars are seen from a distance of several Mpc from the Sun
(\eg in Cen~A, Rejkuba 2004).

Pulsating variables of $\delta$~Sct, $\gamma$~Dor, $\beta$~Cep type,
and Slowly Pulsating B (SPB) stars showing multiple modes are of
particular interest for testing seismic models of MS and
post-MS stars (\eg Daszy\'nska-Daszkiewicz, Dziembowski and Pamyatnykh 2005,
Balona and Dziembowski 2011, Balona \etal 2012). Information on periodicities
permit the determination of basic parameters of stars, such as mass,
effective temperature, and metallicity (\eg Li, Xu and Li 2010,
Breger \etal 2011, Papar\'o \etal 2013). Recently, Su\'arez \etal (2010)
proposed to use rotational splitting asymmetries to probe
the internal rotation profile of $\beta$~Cep type stars.

For more than two decades the number of detected pulsating stars has
increased rapidly. For example, the Optical Gravitational Lensing Experiment
(OGLE) -- a long-term large-scale sky survey focused on stellar
variability (Udalski \etal 1992, Udalski 2003) has already discovered
and classified about 400~000 pulsating variables in the Galactic bulge
and Magellanic Clouds (\eg Soszy\'nski \etal 2010, 2011ab, 2013).

This paper presents results of a search for pulsating stars in twenty-one
OGLE-III fields located in the Galactic disk toward constellations
of Carina, Centaurus, and Musca. In the following sections
we describe: the observations and data reductions (Section~2), the on-line
material (Section~3), the detected variable stars of different types
in details (Sections~4), and the completeness of the search (Section~5).
In Section~6, we summarize our results, while in the last section,
Section~7, we introduce the on-going OGLE-IV Galactic Disk Survey.

%%%%%%%%%%%%%%%%%%%%%%%%%%%%%%%%%%%%%%%%%%%%%%%%%%%%%%%%%%%%%%%%%%

\Section{Observations and Data Reductions}

The observations presented in this paper were collected with the 1.3-m
Warsaw telescope at Las Campanas Observatory, Chile, during the third
phase of the Optical Gravitational Lensing Experiment (OGLE-III)
in years 2001--2009. The observatory is operated by the Carnegie
Institution for Science. The OGLE-III CCD camera consisted of eight
chips with the total field of view of $35\arcm \times 35\arcm$
and the scale of 0.26 arcsec/pixel. More details on the instrumentation
setup can be found in Udalski (2003).

The OGLE-III Galactic disk fields covered an area of 7.12~deg$^2$
near the Galactic plane between longitudes $+288\arcd$ and $+308\arcd$.
Coordinates of the centers, the time coverage, and the number of data
points of the monitored fields are given in Table~1 and shown in Figs.~1--2
in Pietrukowicz \etal (2013). A significant majority of frames (815--2698
per field) were collected in the standard $I$-band filter with an exposure time
of 120 or 180~s. Additional observations, consisting of only 3--8 measurements,
were taken in the $V$-band filter with an exposure time of 240~s.
A total number of approximately $8.8 \times 10^6$ stars with brightness
between $I=12.5$~mag and $I=21.5$~mag was observed (Szyma\'nski \etal 2010).

The photometry was obtained with the standard OGLE data reduction pipeline
(Udalski \etal 2008) using the Difference Image Analysis (Alard and Lupton 1998,
Wo\'zniak 2000). We performed a frequency search up to 24~d$^{-1}$ with the help
of the F{\footnotesize NPEAKS} code (written by Z. Ko{\l}aczkowski, private
communication) for stars containing at least 30 measurements in the $I$-band.
Around 345~500 detections were visually inspected for any kind of variability.
This was done in two stages. In the first stage, we inspected about 100~500
stars having a variability signal-to-noise ratio S/N$\geq14$, and in the second
stage about 245~000 objects with $10\leq$S/N$<14$. We also performed an additional
search for high frequencies between 24~d$^{-1}$ and 100~d$^{-1}$ for stars
with $I<18.5$~mag and $V-I<0.7$~mag. Our searches resulted in thousands
of miscellaneous variables (\eg Mr\'oz \etal 2013, Pietrukowicz \etal 2013,
2014, in preparation).

Based on the initially determined period and shape of the light curve we classified
our candidate pulsating stars into the following types: Classical Cepheids
(20 objects), RR~Lyr type stars (45 objects), Long-Period Variables (31 objects),
and short-period stars for further analysis (about 2000 objects with $P<0.25$~d).
We manually removed obvious outliers from the light curves of our stars
and improved their periods using the TATRY code (Schwarzenberg-Czerny 1996).
In the next step, we searched the Cepheids and short-period stars for additional
modes. Based on the periods, period ratios, and amplitudes of the
short-period variables we selected 57 good candidates for $\delta$~Sct type stars.
Another candidate for a $\delta$~Sct variable together with a very likely pulsating
white dwarf were found as a result of the high-frequency search.

Each variable was calibrated from the instrumental to the standard magnitudes
using transformation relations given in Udalski \etal (2008).
Since the number of $V$-band measurements is very small, we rely
on average instrumental brightnesses $\langle v \rangle$
and $\langle i \rangle$, and color $\langle v \rangle-\langle i \rangle$.
For high-amplitude stars, we fit the instrumental $i$-band light curve to the
instrumental $v$-band data by adjusting the amplitude and phase shift.
The average brightness in both passbands was derived from the fits.
In the case of low-amplitude multi-mode pulsating stars the $\langle v \rangle$
and $\langle i \rangle$ values were determined as the arithmetic
mean from all measurements after converting magnitudes into fluxes.
For stars without any available $V$-band measurement, except LPVs,
we assumed an average $V-I$ color of neighboring stars within a radius
of $1\zdot\arcm0$ taken from the photometric maps in Szyma\'nski \etal (2010).
For LPVs without points in $V$, we assumed an average $V-I$ color determined
from the remaining LPVs. For all stars with $V-I>1.5$~mag the final
$I$-band photometry had to be additionally corrected according to formula
presented in Szyma\'nski \etal (2011). 

%%%%%%%%%%%%%%%%%%%%%%%%%%%%%%%%%%%%%%%%%%%%%%%%%%%%%%%%%%%%%%%%%%%%

\Section{On-line Data}

The catalog of pulsating stars in the Galactic disk fields,
containing tables with basic parameters, time-series $I$- and $V$-band
photometry, and finding charts, is available to the astronomical
community from the OGLE Internet Archive:
\begin{center}
{\it http://ogle.astrouw.edu.pl\\}
\end{center}
and
\begin{center}
{\it ftp://ftp.astrouw.edu.pl/ogle/ogle3/OIII-CVS/gd/\\}
\end{center}
where one can find the following catalogs on pulsating stars: {\sf cep},
{\sf dsct}, {\sf lpv}, {\sf rrlyr}, {\sf t2cep}, {\sf wd}, and {\sf unclassified}.
The stars are arranged according to increasing right ascension and named as
OGLE-GD-CEP-NNNN, OGLE-GD-DSCT-NNNN, OGLE-GD-LPV-NNNN, OGLE-GD-RRLYR-NNNN,
OGLE-GD-T2CEP-NNNN, and OGLE-GD-WD-NNNN, where NNNN are four-digit consecutive
numbers for each of the type. In the case of short-period unclassified
variables we leave the name as the following catenation: FIELD.CHIP.ID.
In the data tables, we provide coordinates of the
variables, their pulsation periods, period uncertainties, information
on brightness, and classification to a subtype, if exists.

%%%%%%%%%%%%%%%%%%%%%%%%%%%%%%%%%%%%%%%%%%%%%%%%%%%%%%%%%%%%%%%%%%%%%

\Section{Detected Pulsating Variables}

\Subsection{Classical Cepheids}

Our search for variable objects brings identification of 20 Classical
Cepheids. All Cepheids but one object are new discoveries.
Variable OGLE-GD-CEP-0019 was detected by the EROS-II
collaboration during the Galactic Spiral Arms (GSA) observation
program and named EROS2 GSA J133319-640707 (Derue \etal 2002).
Light curves of all OGLE-III disk Classical Cepheids are presented in Fig.~1.
Eight stars are double-mode pulsators (beat Cepheids) of which three
pulsate in the fundamental mode (F) and first overtone (1O), four
in the first and second overtone (2O), and one object, OGLE-GD-CEP-0001,
in the first overtone and a second mode of unknown origin (labeled with X).
Ten Cepheids pulsate in the fundamental mode. The pulsation period of
OGLE-GD-CEP-0010 is longer than 10~d, hence this star
is younger than 50 Myr (based on models in Bono \etal 2005).
From the information on the period ratio for F/1O pulsators, we infer
that OGLE-GD-CEP-0012 and OGLE-GD-CEP-0016 are very likely objects
of lower metallicity in comparison to other stars of this type
(see Fig.~10 and discussion in Section 4.5).

\begin{figure}[htb!]
\centerline{\includegraphics[angle=0,width=130mm]{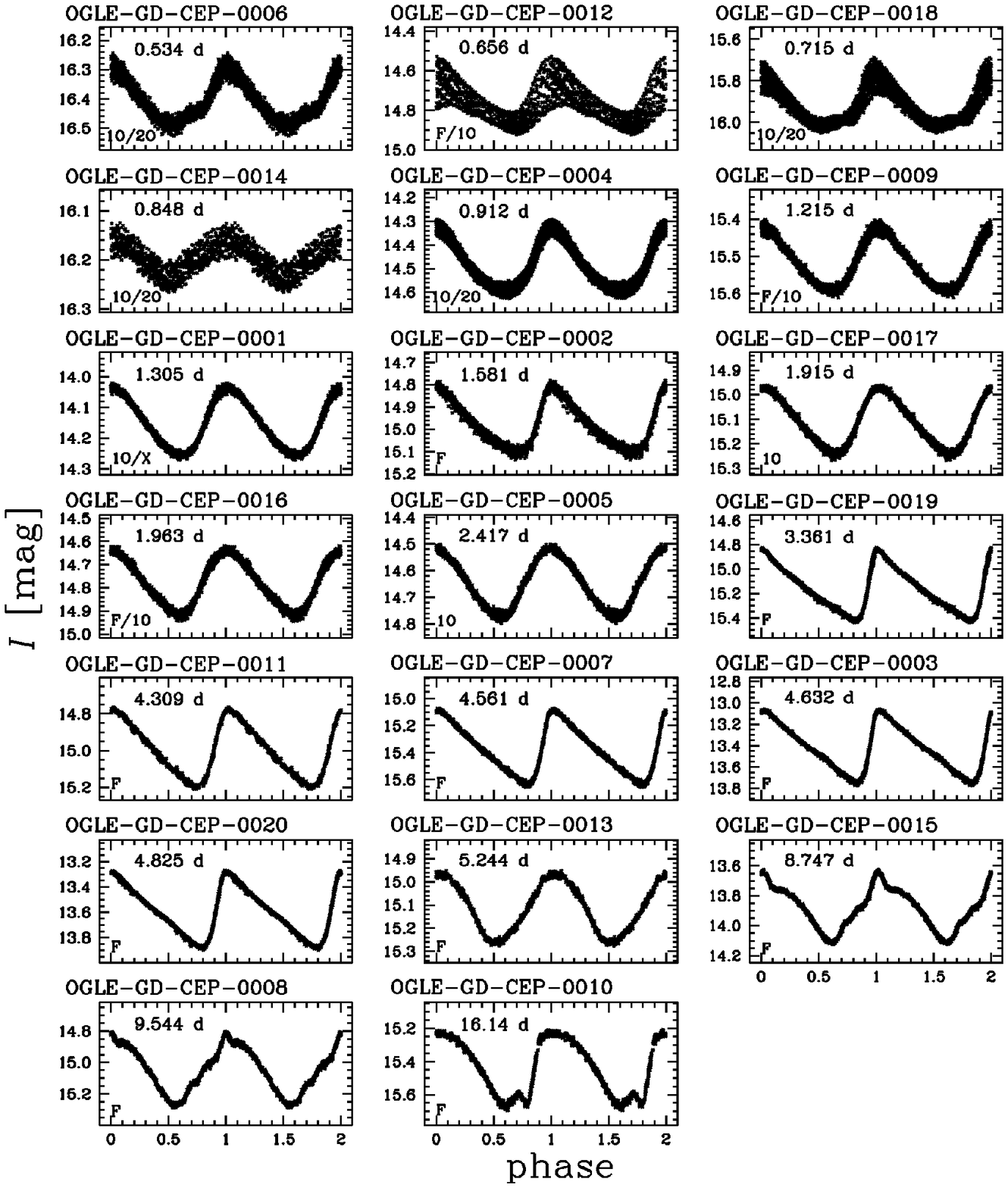}}
\FigCap{Phased light curves of 20 Classical Cepheids detected in the
OGLE-III disk area. Objects are sorted according to increasing period.}
\end{figure}

All Cepheids have the average apparent brightness between $I=13.4$~mag
and $I=16.4$~mag. Star OGLE-GD-CEP-0011 has no measurement in $V$ due
to unfortunate position just outside the reference image in this band.
The observed $V-I$ color of the remaining 19 Classical Cepheids
is in a range between 1.1~mag and 4.2~mag. Such a wide magnitude range
is the result of the non-uniformly distributed interstellar extinction
in the Galactic disk. It is worth to note that the monitored area
contains other eleven known Classical Cepheids, but all of them are
saturated in the OGLE data.

%%%%%%%%%%%%%%%%%%%%%%%%%%%%%%%%%%%%%%%%%%%%%%%%%%%%%%%%%%%%%%%%%%%%%

\Subsection{Candidates for Type II Cepheids}

The observed amplitudes (0.2--0.4~mag) and shape of the light curves
of six stars (Fig.~2) resemble those of Type II Cepheids (\eg Soszynski
\etal 2011b). Five stars with periods between 5~d and 20~d could be of
W~Vir type, while the 35.15~d star OGLE-GD-T2CEP-0004 is a probable RV~Tau
type variable. For object OGLE-GD-T2CEP-0001 there is no color information
due to heavy reddening, while the remaining stars have $V-I$ between
$1.9$ and $2.7$~mag. Their location in the $I$ \vs $V-I$ diagrams
(Fig.~3) is in agreement with the proposed types of old population
Cepheids. However, our confidence is not full due to the fact that
all objects but OGLE-GD-T2CEP-0001 were detected at an angular distance
$<0\zdot\arcd8$ from the Galactic plane in fields
observed for about one season only.

\begin{figure}[htb!]
\centerline{\includegraphics[angle=0,width=130mm]{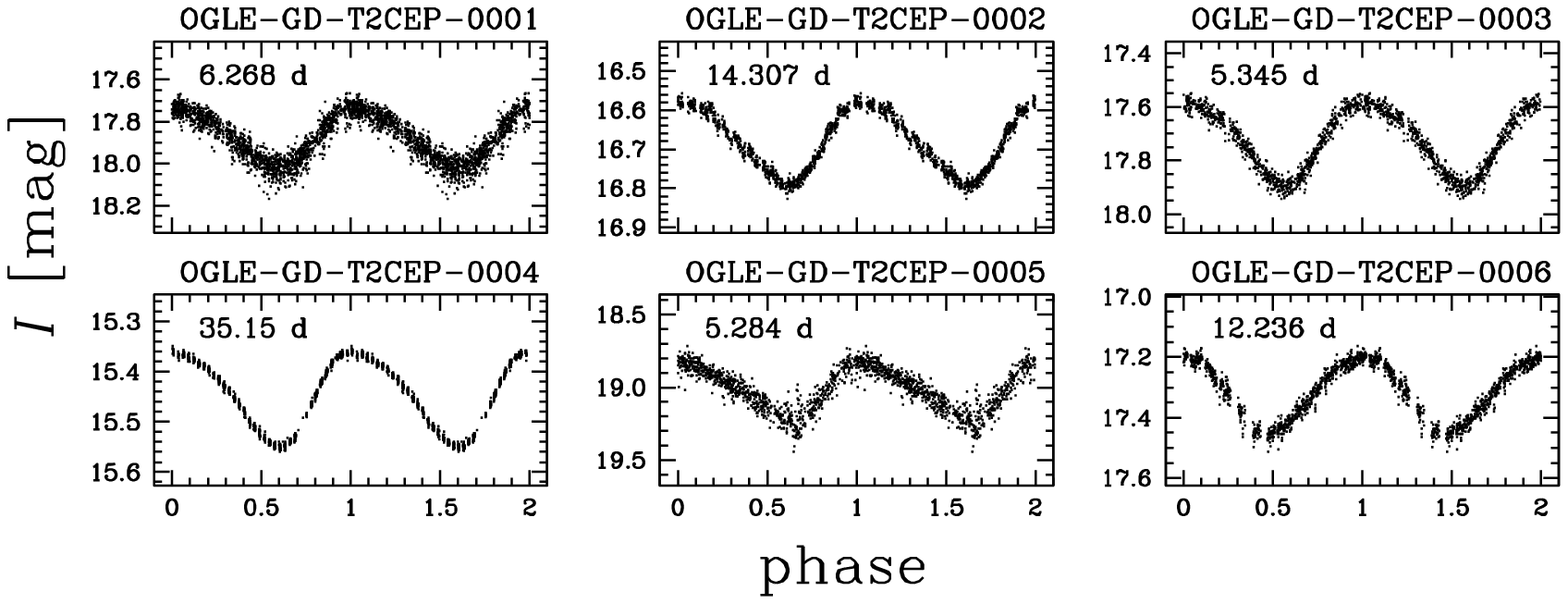}}
\FigCap{Phased light curves of six candidates for Type II Cepheids.}
\end{figure}

\begin{figure}[htb!]
\centerline{\includegraphics[angle=0,width=130mm]{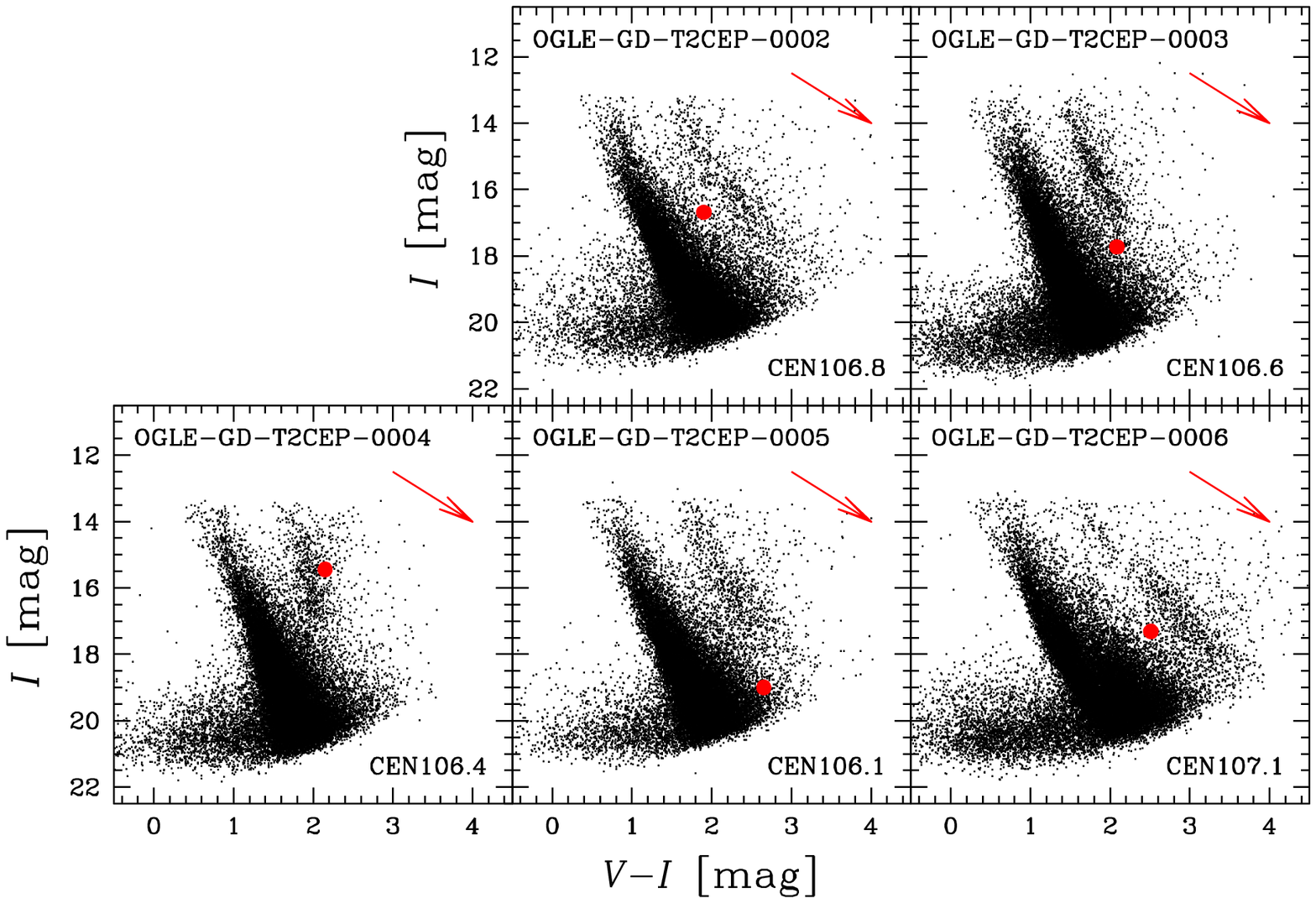}}
\FigCap{Color-magnitude diagrams with marked positions (red points)
of candidates for Type II Cepheids. Field name is given in each panel.
Red arrows indicate the direction of reddening with an assumed
ratio of total-to-selective exinction $R_I=A_I/E(V-I)=1.5$.}
\end{figure}

%%%%%%%%%%%%%%%%%%%%%%%%%%%%%%%%%%%%%%%%%%%%%%%%%%%%%%%%%%%%%%%%%%%%%

\Subsection{RR~Lyr type Stars}

Within the twenty-one disk fields we found 45 RR~Lyr stars of which 36 are
of the type ``ab'' (fundamental mode) and nine of the type ``c'' (first overtone).
Light curves of the variables are presented in Fig.~4. Object OGLE-GD-RRLYR-0005
was detected independently in two overlapping fields CAR109 and CAR115.
The increased time span allowed a better determination of the pulsation
period for this star. Twelve of our RRab variables (\ie one-third of the sample)
exhibit the Blazhko effect (Bla\v{z}ko 1907), which manifests as long-period
modulations in the light curve shape (Smith 2004). The observed Blazhko
periods are between 21~d and 90~d.

\begin{figure}[htb!]
\centerline{\includegraphics[angle=0,width=130mm]{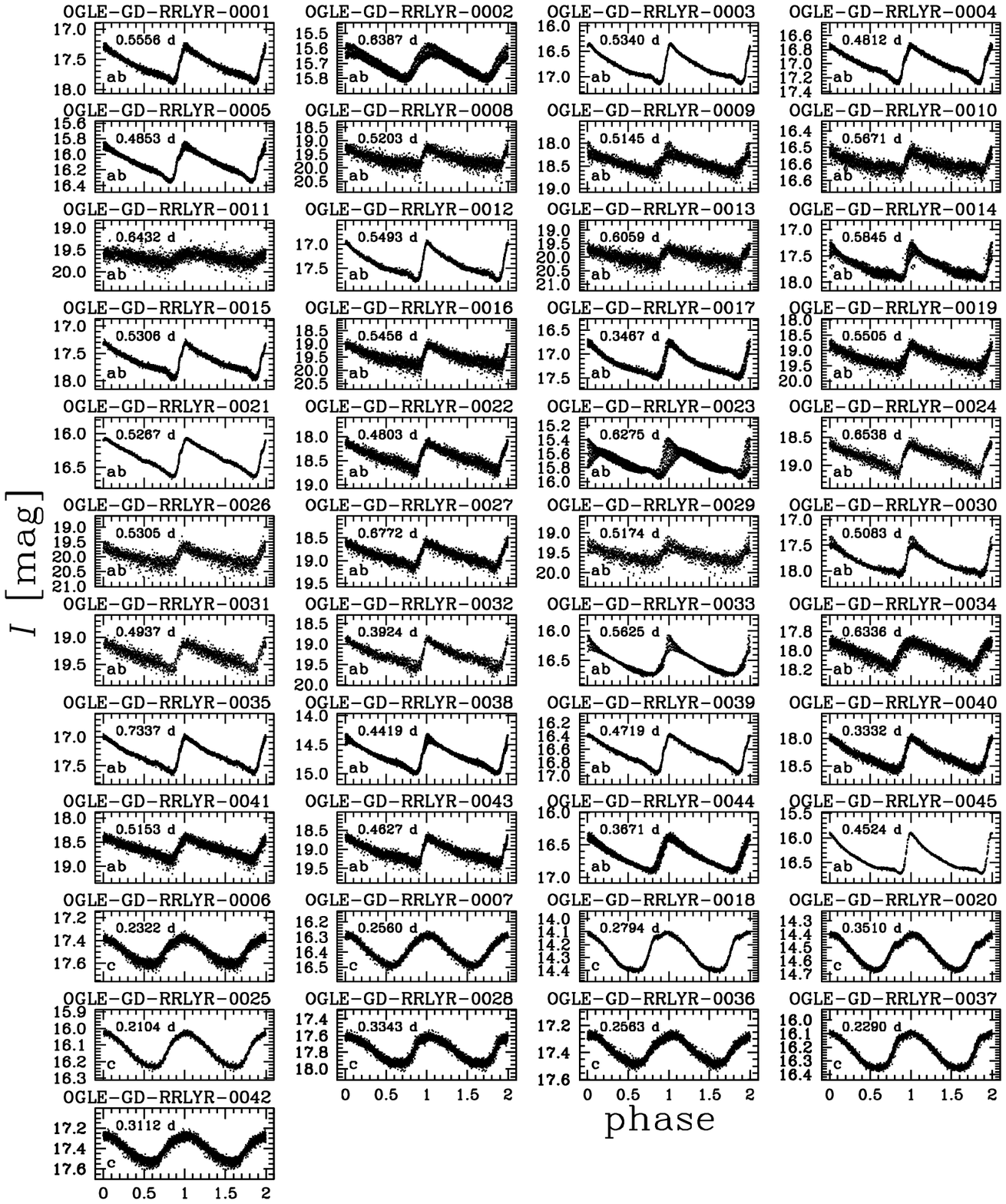}}
\FigCap{Phased light curves of 45 RR~Lyr variables found within
the OGLE-III Galactic disk fields. The type of the variables
is given in the lower left corner of each panel.}
\end{figure}

We used an empirical relation presented in Smolec (2005) to assess
the metallicity of the RRab stars on the Jurcsik (1995) scale (J95):
\begin{displaymath}
{\rm \feh_{J95}} = - 3.142 - 4.902 P + 0.824 \phi_{31},~~\sigma_{\rm sys}=0.18,
\end{displaymath}
where $P$ is the pulsation period and $\phi_{31} = \phi_{3}-3\phi_{1}$
is a Fourier phase combination derived from the $I$-band light curve.
In Fig.~5, we present the obtained metallicity distributions
for all RRab stars as well as those without the Blazhko effect.
The stars form essentially two groups: a metal-rich one with
[Fe/H]$_{\rm J95}\gtrsim-0.8$ and a metal-poor one with
[Fe/H]$_{\rm J95}\lesssim-0.8$. This division likely reflects
the presence of stars from two Galactic populations: the thick disk
and halo, respectively. Such explanation is in agreement with
properties of RR~Lyr stars observed in the Galaxy (\eg Kinemuchi \etal 2006,
Kinman, Morrison and Brown 2009, Szczygie{\l} \etal 2009).

\begin{figure}[htb!]
\centerline{\includegraphics[angle=0,width=110mm]{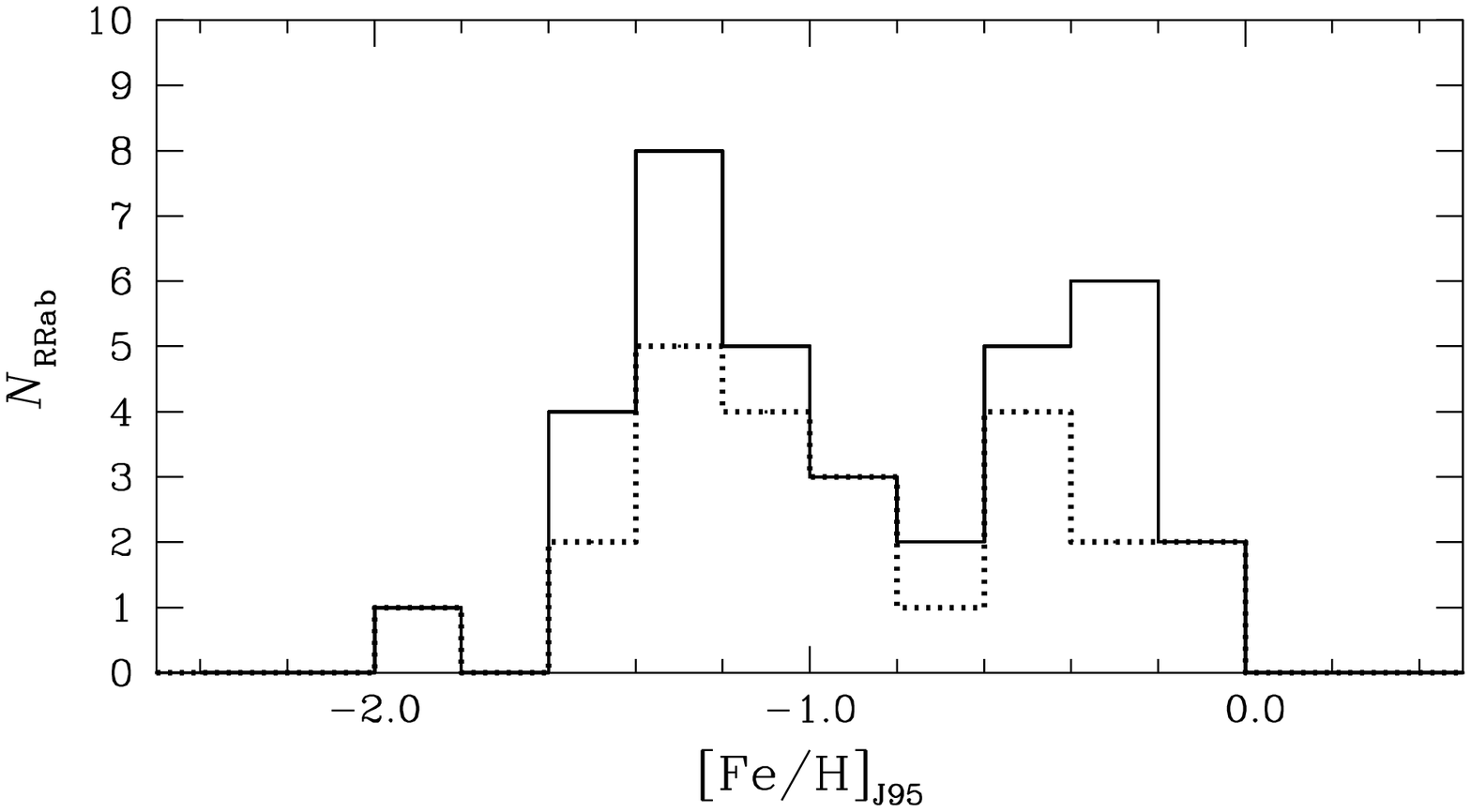}}
\FigCap{Distributions of photometrically derived metallicities for all
36 newly detected RRab type stars (solid line) and a subgroup of 24 variables
without the Blazhko effect (dotted line). The metallicity is given
in the Jurcsik (1995) scale.}
\end{figure}

%%%%%%%%%%%%%%%%%%%%%%%%%%%%%%%%%%%%%%%%%%%%%%%%%%%%%%%%%%%%%%%%%%%%%

\Subsection{Long-Period Variables}

We identified 31 new LPVs in the OGLE-III Galactic disk area.
In Fig.~6, we present their $I$-band light curves phased with our
best-measured pulsation periods. Eighteen of the LPVs are Mira stars,
while the other thirteen objects are Semi-Regular Variable (SRV) stars.
For ten LPVs the observations are insufficient to estimate their
peak to peak amplitudes in $I$. Due to the lack of $V$-band data
we were not able to assess the $V-I$ color for four Mira stars.
Miras OGLE-GD-LPV-0023 and 0027 have the largest mean observed color among
all detected variables within the OGLE-III disk fields, $V-I\approx7.6$~mag.

\begin{figure}[htb!]
\centerline{\includegraphics[angle=0,width=130mm]{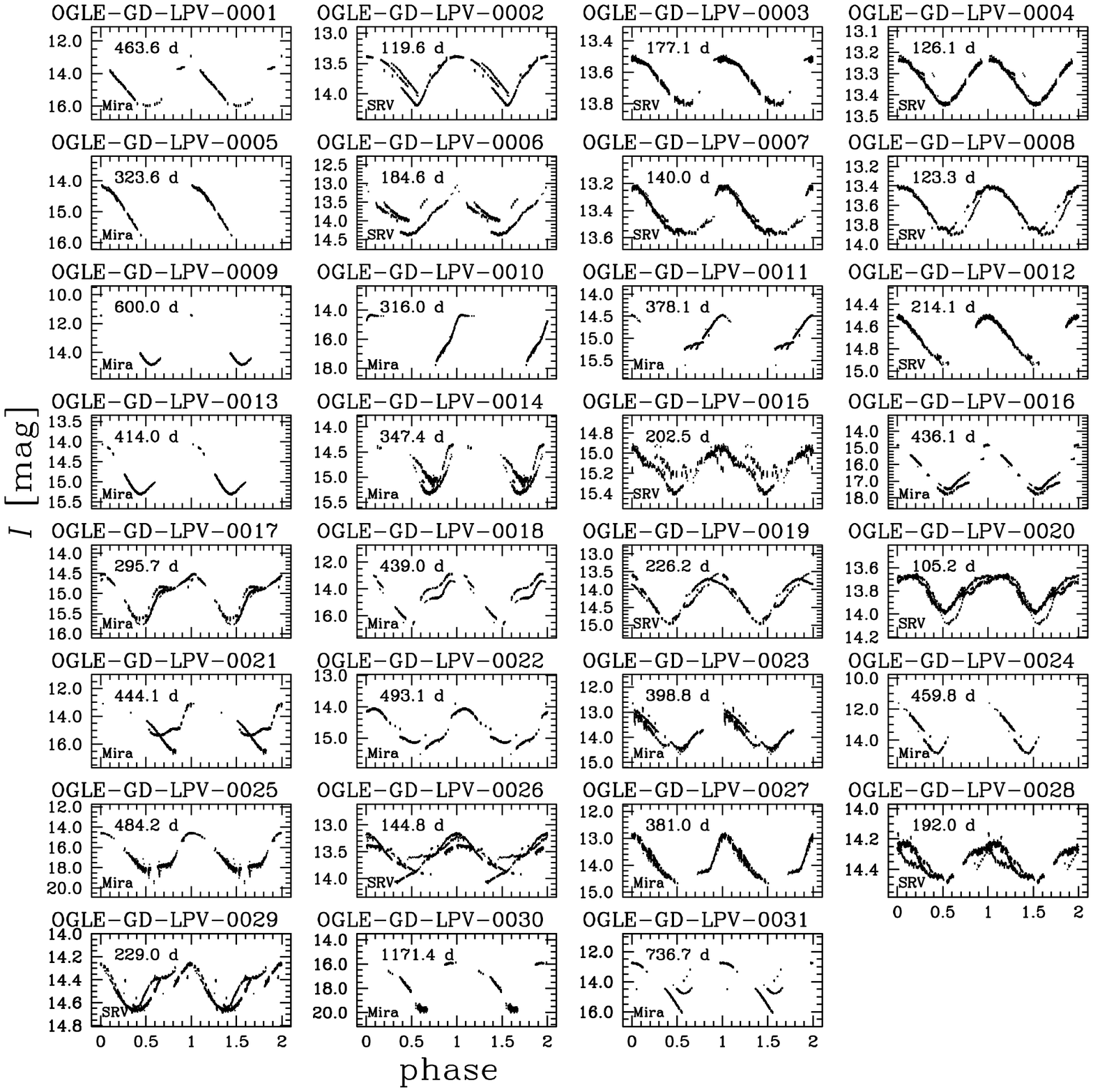}}
\FigCap{Light curves of 31 LPVs detected in the OGLE-III disk area.}
\end{figure}

%%%%%%%%%%%%%%%%%%%%%%%%%%%%%%%%%%%%%%%%%%%%%%%%%%%%%%%%%%%%%%%%%%%%%

\Subsection{$\delta$~Sct type Stars}

We classified 58 stars as very likely $\delta$~Sct type variables.
Fifty-seven of them were found during the search
for frequencies lower than 24~d$^{-1}$ and one object during
the high-frequency search (24--100~d$^{-1}$). The alternative within
the realm of short-period pulsating stars is the $\beta$~Cep type.
The period ranges for these two types largely overlap, although $\beta$~Cep
stars have much higher masses and effective temperatures. Unfortunately,
our data do not allow us to assess these parameters, mostly because of
the heavy and unknown reddening in the observed regions.
Thus, we cannot exclude that some of the stars, which we classified as
$\delta$~Sct, may be in fact $\beta$~Cep stars. However, this seems
unlikely because amplitudes as high as in our stars ($>0.1$~mag)
are rare in $\beta$~Cep stars (see Stankov \& Handler 2005).
Furthermore, the period range in our multi-periodic objects is wider
than typically found in $\beta$~Cep stars.

Twenty-eight stars in our sample are fundamental-mode pulsators
(see light curves in Fig.~7). Of course we cannot exclude
that some other modes are excited in these objects, but their
amplitudes must be much smaller. Stars with one or two dominant
radial modes are known as High Amplitude Delta Scuti stars (HADS).
They are rare. According to Lee \etal (2008) only about 0.3\% of
the total population of Galactic $\delta$~Sct stars belongs to this
subtype. What causes that a star chooses this high-amplitude
form of pulsation is unknown.

\begin{figure}[htb!]
\centerline{\includegraphics[angle=0,width=130mm]{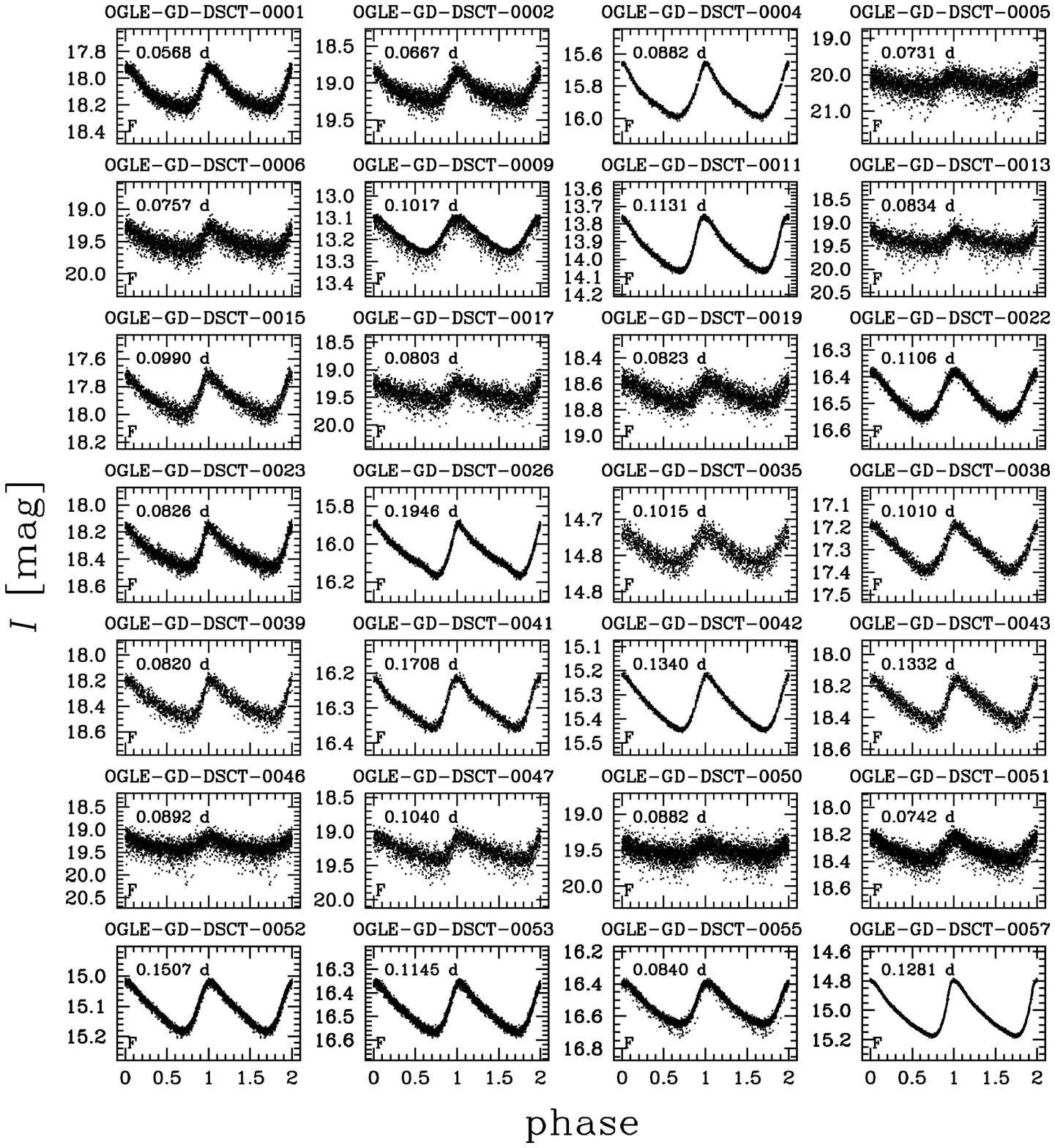}}
\FigCap{Phased light curves of 28 detected fundamental-mode $\delta$~Sct
type variables.}
\end{figure}

In the remaining 30 stars of our sample we detected at least two modes.
Their light curves phased with the strongest period are plotted
in Fig.~8. The amplitudes of light variations are large
by $\delta$~Sct star standards. However, it is not clear if the
dominant peaks are due to radial modes. In this work, we assume this
to be the case. Only with this optimistic assumption an inference
on stellar parameters based on periods alone is possible.

\begin{figure}[htb!]
\centerline{\includegraphics[angle=0,width=130mm]{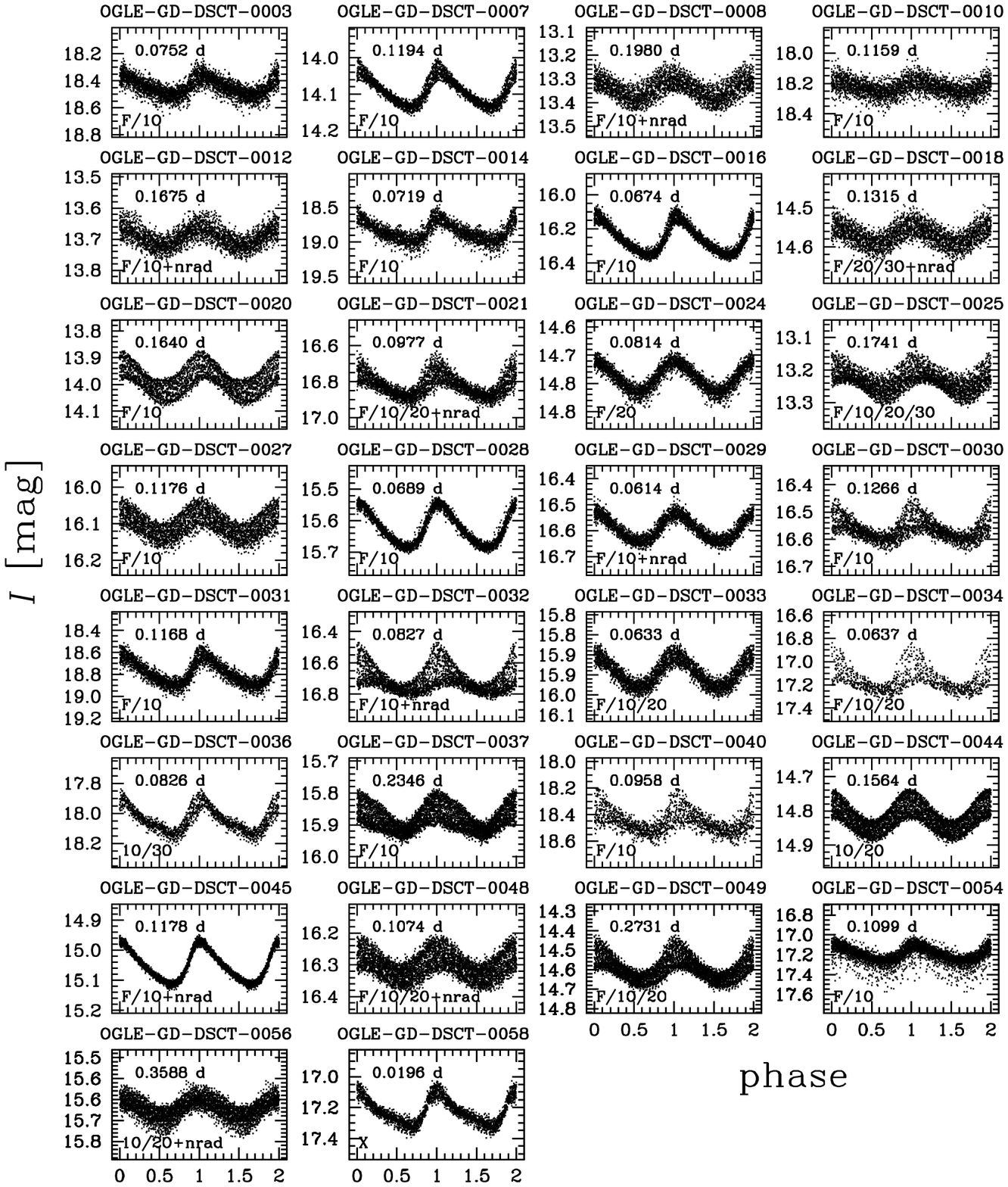}}
\FigCap{Light curves of 30 detected multi-mode $\delta$~Sct type stars.}
\end{figure}

Twenty-two $\delta$~Sct stars exhibit two radial modes (in eighteen cases
F and 1O), six stars exhibit three radial modes (mostly F, 1O, and
2O), and one star, OGLE-GD-DSCT-0025, seems to have four radial
modes (F, 1O, 2O, and 3O, see periodograms in Fig.~9). In some of
the stars, we detected additional non-radial modes. Objects
with more than two {\it bona fide} radial modes are particularly
valuable in this context. Table~1 lists data for such objects.

\begin{figure}[htb!]
\centerline{\includegraphics[angle=0,width=130mm]{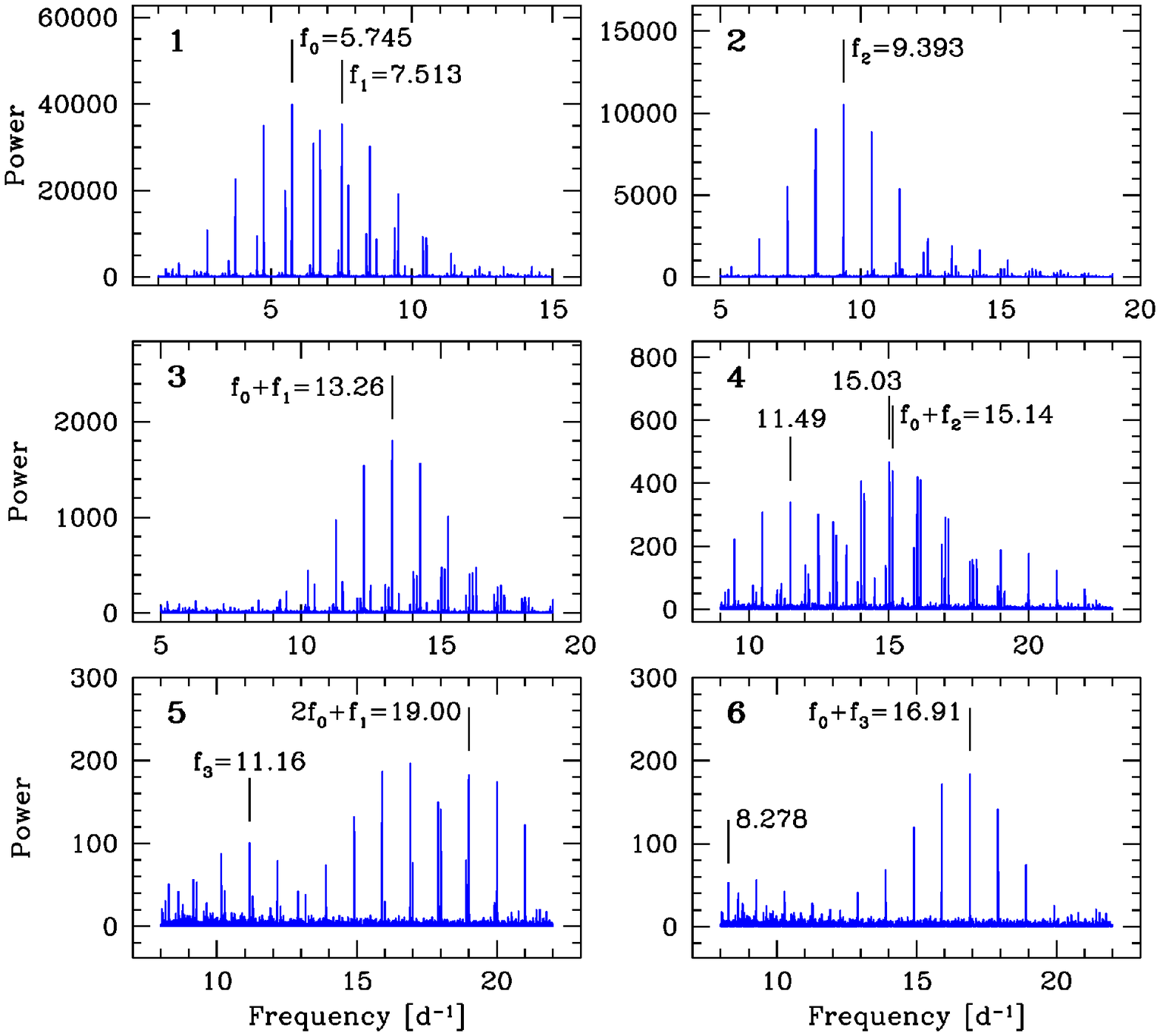}}
\FigCap{Periodograms with marked frequencies after subsequent
pre-whitenings (numbered in the upper left corners) for the
quadruple-mode star OGLE-GD-DSCT-0025. Radial modes and their
combinations are labeled.}
\end{figure}

The identification of the radial orders of excited modes in
multi-mode pulsators was done with the help of the Petersen diagram
(shorter-to-longer period ratio \vs the logarithm of the longer
period). The diagram, presented in Fig.~10, includes the multi-mode
$\delta$~Sct as well as double-mode $\delta$~Cep type stars. The
relations for the former type looks like continuations of the
relations for multi-mode Cepheids. This has been already noted for
numerous variables detected in the Large Magellanic Cloud (see
Fig.~3 in Poleski \etal 2010) and is not surprising because
post-MS $\delta$~Sct stars are just a low-mass version
Cepheids crossing the instability strip for the first time. What
matters is that this continuity supports the radial mode hypothesis
for modes in $\delta$~Sct stars.

One of the disk Classical Cepheids, OGLE-GD-CEP-0001, is a first-overtone
pulsator with an additional secondary period near $0.6P_{1\rm O}$.
About 140 such Cepheids from the Small Magellanic Cloud (SMC)
and 30 from the Large Magellanic Cloud (LMC) were reported
in Soszy\'nski \etal (2008) and Soszy\'nski \etal (2010), respectively.
The origin of this secondary period remains unknown.

\begin{figure}[htb!]
\centerline{\includegraphics[angle=0,width=130mm]{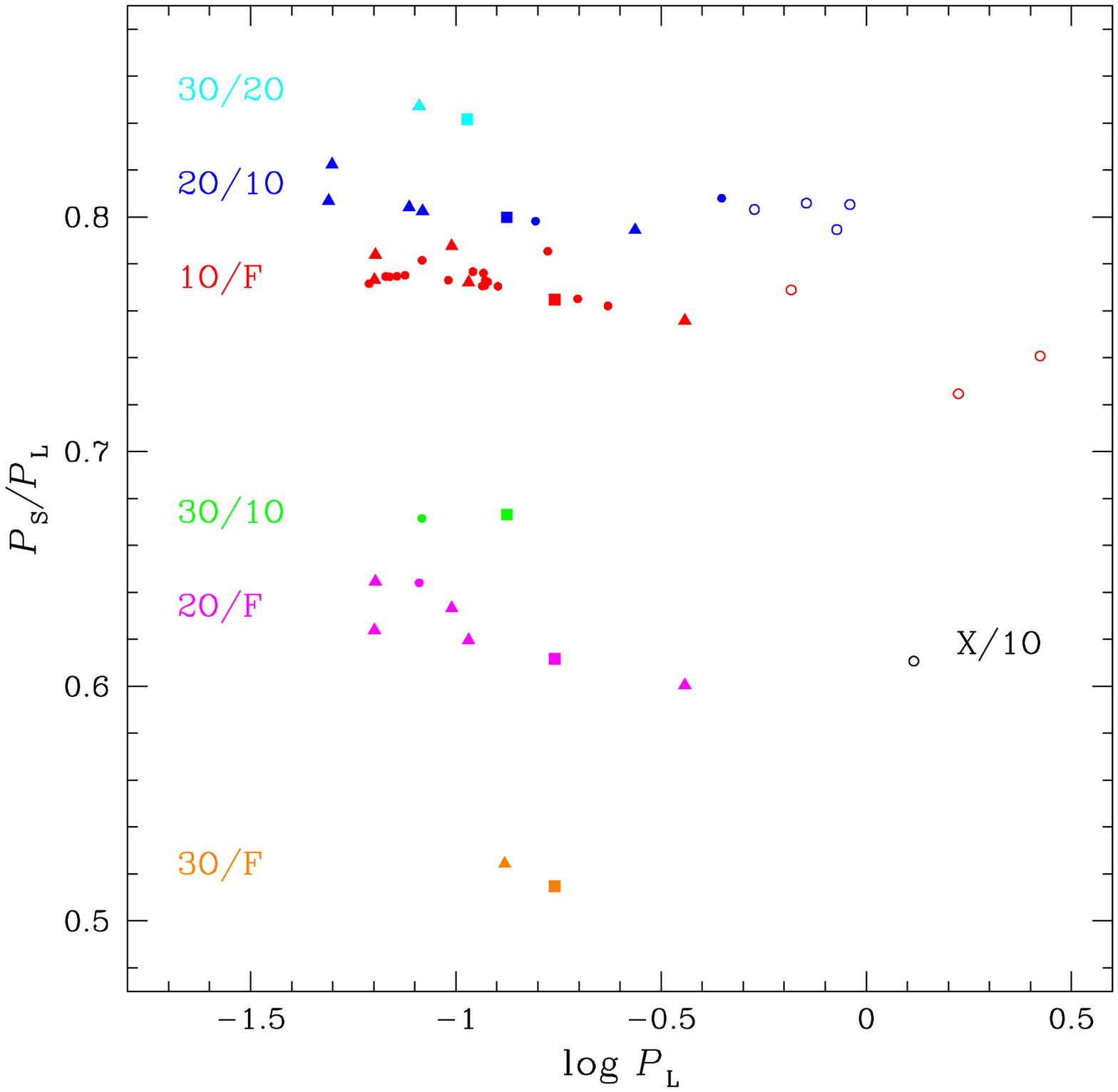}}
\FigCap{Petersen diagram for multi-mode pulsators in the OGLE-III disk area.
Open symbols correspond to Classical Cepheids, filled symbols to
$\delta$~Sct type stars. Circles refer to double-mode, triangles
-- triple-mode, and squares -- quadruple-mode pulsators.}
\end{figure}

\begin{table}[htb!]
\centering
\caption{\small Observed periods for seven $\delta$~Sct type stars
with at least three radial modes}
\medskip
{\small
\begin{tabular}{ccccc}
\hline
Variable      &   $P_{\rm F}$   &  $P_{\rm 1O}$  &  $P_{\rm 2O}$  &  $P_{\rm 3O}$ \\
OGLE-GD-      &                 &                &                & \\
-DSCT-        &      [d]        &       [d]      &       [d]      &      [d] \\
\hline
\hline
-0018         & 0.13152629(81)  &        -       & 0.08140070(20) & 0.06896097(53) \\
-0021         & 0.09766834(39)  & 0.07692128(44) & 0.06186007(56) & - \\
-0025         & 0.1740594(19)   & 0.13310790(70) & 0.10646403(51) & 0.0896032(29) \\
-0033         & 0.06330743(12)  & 0.04894428(14) & 0.03949024(70) & - \\
-0034         & 0.06369610(25)  & 0.04993199(9)  & 0.04106100(26) & - \\
-0048         & 0.10740182(5)   & 0.08293446(1)  & 0.06655117(3)  & - \\
-0049         & 0.36141456(19)  & 0.27314926(19) & 0.21699323(26) & - \\
\hline
\end{tabular}}
\end{table}

An interesting application of the Petersen diagrams is probing of
metal abundance in stars (see Popielski, Dziembowski and Cassisi 2000,
Buchler and Szab\'o 2007, Buchler 2008, and Soszy\'nski \etal 2011a for
application to RR~Lyr stars and Cepheids). We applied this tool to
27 $\delta$~Sct stars of our sample, which reveal pairs of radial
modes of consecutive orders. The data are compared with model values
in the separate diagrams employing the (F,1O) and (1O,2O) pairs
shown in Fig.~11. The lines present the values calculated for stellar
models along evolutionary tracks at selected metallicity parameters,
$Z$, and masses, $M$, which are depicted in Fig.~11.
The tracks extend over the MS (in some cases) and post-MS part
of the pulsation instability range. The relative hydrogen abundance
$X=0.72$ and the mixing length parameter $\alpha_{\rm MLT}=1.5$
were adopted in all these models. Effects of overshooting and
rotation have been ignored but this choice is not essential here.

\begin{figure}[htb!]
\epsfxsize=1\hsize\epsffile{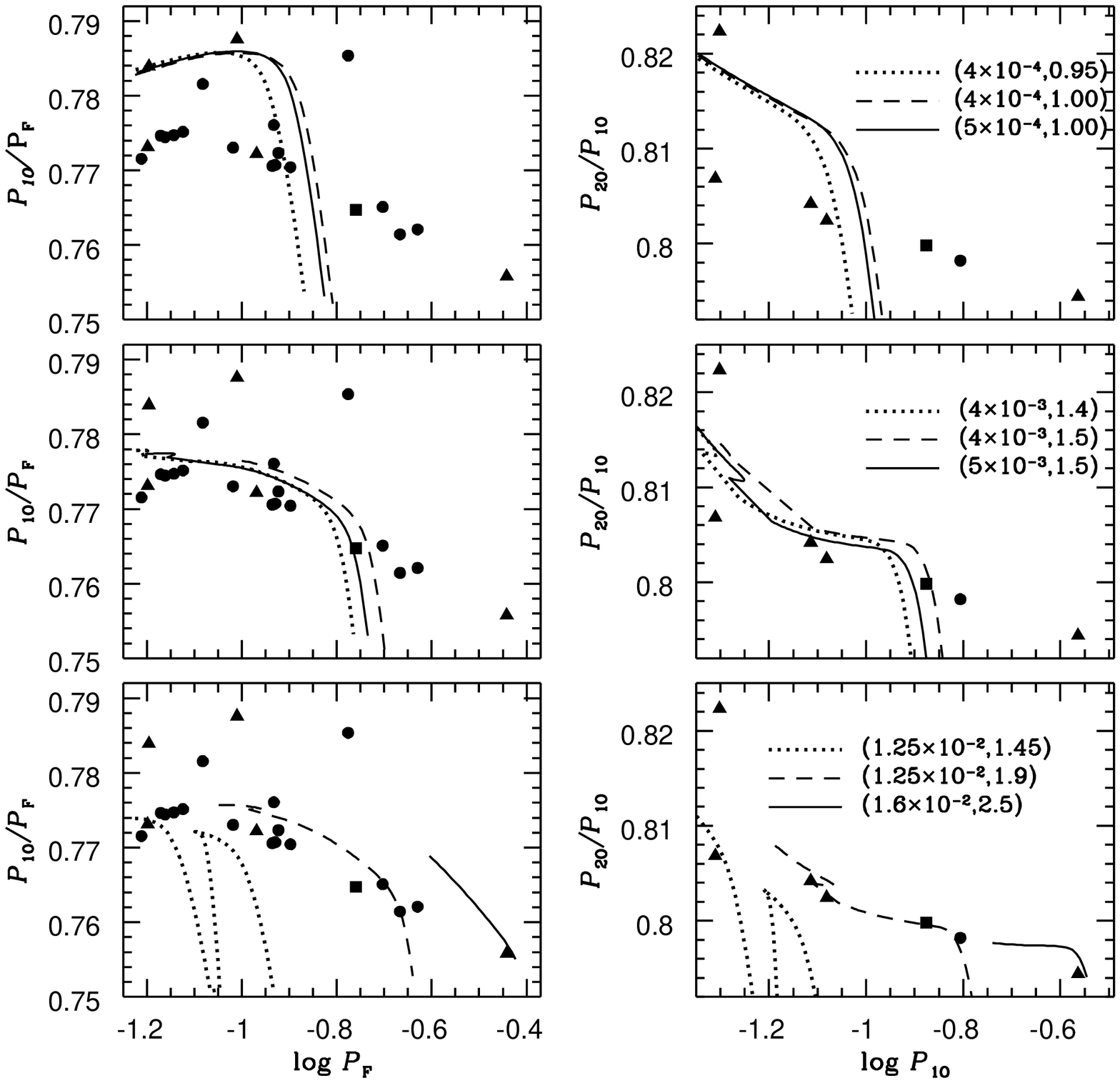}
\caption{Location in the Petersen diagrams of objects with more
than one {\it bona fide} radial mode detected compared with the
model predictions. The left and right panels refer to the (F,1O) and
(1O,2O) pairs, respectively. Circles, triangles, and squares
refer to stars with two, three, and four radial modes detected, respectively.
The same evolutionary models were used in both panels of each row.
The numbers shown in the right panels give the metallicity $Z$
and mass in $M_{\odot}$.}
\end{figure}

We may see that a large range of $Z$-values is required to explain
star positions in both diagrams. For triple-mode stars OGLE-GD-DSCT-0021
and 0034 in Table~1 we need $Z\approx4\times10^{-4}$, which corresponds
to Population II objects, called SX~Phe stars. Similar is required
for double-mode object OGLE-GD-DSCT-0032 at $\log P_{\rm F}\approx-1.08$
and $P_{1\rm O}/P_{\rm F}\approx0.782$. There is one double-mode outlier,
0012, at $\log P_{\rm F}\approx-0.78$ and $P_{1\rm O}/P_{\rm F}\approx0.785$,
which certainly cannot be reproduced by our models.
It is worth noting that $\beta$~Cep stars, which are much more massive
and metal rich may populate this part of the diagram. To explain
positions of stars OGLE-GD-DSCT-0025, 0033, 0048, and 0049 and a number
of double-mode objects, a close to solar metal abundance is needed.

With data on three radial mode periods (Table~1), adopting the same conditions
as for calculations made for the models used in Fig.~11., we may try to
determine exactly all stellar parameters. Proceeding this way,
Moskalik \& Dziembowski (2005) calculated their seismic models of
two triple-mode Cepheids. We tried the same method and found that it had
worked well for some of our stars but not all of them. The problem may be
seen in the upper panels of Fig.~11. Consider for instance star OGLE-GD-DSCT-0021
($\log P_{\rm F}\approx-1.01$ and $P_{1\rm O}/P_{\rm F}\approx0.788$).
It is clear that there is no solution for $M$ and $Z$, though we are close
to models with lowest values (solid lines). We found the same for
different choices of $X$, $\alpha_{\rm MLT}$, and  $\alpha_{\rm OV}$
parameters and at moderate rotation. The limited precision of our
period calculations in some cases prohibited the use of this method.
Thus, as the best models of all seven objects in Table~1 were chosen those
minimizing $D\equiv\sqrt{\sum_k(\log P^{\rm obs}_k-\log P^{\rm mod}_k)^2}$.
Adding the 3O period for the quadruple-mode star OGLE-GD-DSCT-0025 to that fit
has a negligible effect on inferred parameters, resulting in the worse
fit and the value of $D$ going up from 0.002 to 0.005.

In Table~2, along with model parameters, we list the reddening,
$E(V-I)$, and distance modulus $m-M$. These quantities were
determined from the mean magnitudes measured in the two bands and
the corresponding absolute magnitudes interpolated from the tables
provided by Castelli and Kurucz (2004). The obtained distances
range from 2.5 to 6.2~kpc, which places the seven variables
in the foreground of stars from the Centaurus spiral arm.

\begin{table}[htb!]
\centering \caption{\small Parameters of our best models for seven
$\delta$~Sct type stars with at least three radial modes}
\medskip
{\small
\begin{tabular}{cccccccc}
\hline
Variable      &     $M$       &   $Z$   &  Age  & log $T_{\rm eff}$ & $E(V-I)$ & $m-M$ & Remarks\\
OGLE-GD-      &               &         &       &                   &          &       & \\
-DSCT-        & [M$_{\odot}$] &         & [Gyr] &        [K]        &   [mag]  & [mag] & \\
\hline \hline
-0018         &     1.74      & 0.00848 &  1.16 &      3.862        &   0.37   & 13.00 & post-MS\\
-0021         &     0.94      & 0.00046 &  5.68 &      3.825        &   0.65   & 13.66 & SX Phe\\
-0025         &     1.92      & 0.01234 &  1.02 &      3.835        &   0.36   & 12.00 & post-MS\\
-0033         &     1.45      & 0.01253 &  1.51 &      3.835        &   0.43   & 12.94 & MS\\
-0034         &     1.01      & 0.00044 &  4.36 &      3.887        &   0.90   & 13.54 & SX Phe\\
-0048         &     1.46      & 0.00713 &  1.83 &      3.842        &   0.53   & 13.95 & post-MS\\
-0049         &     2.46      & 0.01618 &  0.58 &      3.789        &   0.58   & 13.85 & post-MS\\
\hline
\end{tabular}}
\end{table}

The main source of uncertainty in the numbers given in Table~2
is the lack of data on the relative hydrogen abundance and
rotation rate as well as insufficient knowledge about efficiency of
convective transport and overshooting. In order to assess
the range of uncertainties, we carried out calculations with
varing values of $X$, $\alpha_{\rm MLT}$, $\alpha_{\rm OV}$, and
$v_{\rm rot}$. In particular for star OGLE-GD-DSCT-0049, we obtained
the following expressions for inferred parameters with an explicit
dependence on these quantities:
\begin{eqnarray*}
M & = & 2.46-3.33(X-0.72)+0.5(\alpha_{\rm MLT}-1.5)-0.07\alpha_{\rm OV}-0.06\alpha_{\rm rot},\\
Z & = & 0.016-0.27(X-0.72)+0.11(\alpha_{\rm MLT}-1.5)-0.0015\alpha_{\rm OV}-0.0006\alpha_{\rm rot},\\
{\rm Age} & = & 0.58-0.4(X-0.72)+0.15(\alpha_{\rm MLT}-1.5)-0.19\alpha_{\rm OV}-0.034\alpha_{\rm rot},\\
E(V-I) & = & 0.058-0.8(X-0.72)+0.16(\alpha_{\rm MLT}-1.5)-0.19\alpha_{\rm OV}-0.034\alpha_{\rm rot},\\
m-M & =& 13.85-0.8(X-0.72)+0.23(\alpha_{\rm MLT}-1.5)+0.15\alpha_{\rm OV}-0.060\alpha_{\rm rot},\\
\end{eqnarray*}
where
\begin{displaymath}
\alpha_{\rm rot}=\left({v_{\rm rot}\over90~{\rm km/s}}\right)^2.
\end{displaymath}

Searching for frequencies between 24~d$^{-1}$ and 100~d$^{-1}$ led us
to the detection of a star with the dominant period of
0.01962154(1)~d$=28.255018(1)$~min and an exceptionally high
peak-to-peak $I$-band amplitude of 0.35~mag. Its light curve
is presented in the last panel of Fig.~8. The object shows
two additional peaks equally distant from the dominant mode in the
frequency space: $\pm1.9028$~d$^{-1}$ from 50.9644~d$^{-1}$ (Fig.~12).
The high amplitude at the very short pulsation period
causes a problem with proper classification of this object.
It seems to be an extreme case of $\delta$~Sct type stars.
In the revised catalog of 636 $\delta$~Sct stars prepared by
Rodriguez \etal (2000), nineteen objects have periods below
or around 0.03~d, and only three objects below or around 0.02~d. 
The corresponding maximum $V$-band amplitudes are 0.06~mag and 0.02~mag,
respectively. A similar triple-peak structure of the oscillation
spectrum was found in the $\delta$~Sct star 1~Mon (V474 Mon):
$7.3461\pm0.1291$~d$^{-1}$. Balona and Stobie (1980) interpreted 
that structure as a dipolar ($\ell=1$) triplet, where the dominant
peak is due to a radial mode and the two side oscillations are
likely dipole modes split by stellar rotation. If the three peaks
in our star are due to a dipolar triplet then
$\nu_{\rm rot}\approx1.9$~d$^{-1}$, which implies
$v_{\rm rot}\sim100R/R_{\odot}\approx150$~km/s, well acceptable value
for a $\delta$~Sct type object. Therefore we named this intriguing
object as OGLE-GD-DSCT-0058.

\begin{figure}[htb!]
\centerline{\includegraphics[angle=0,width=130mm]{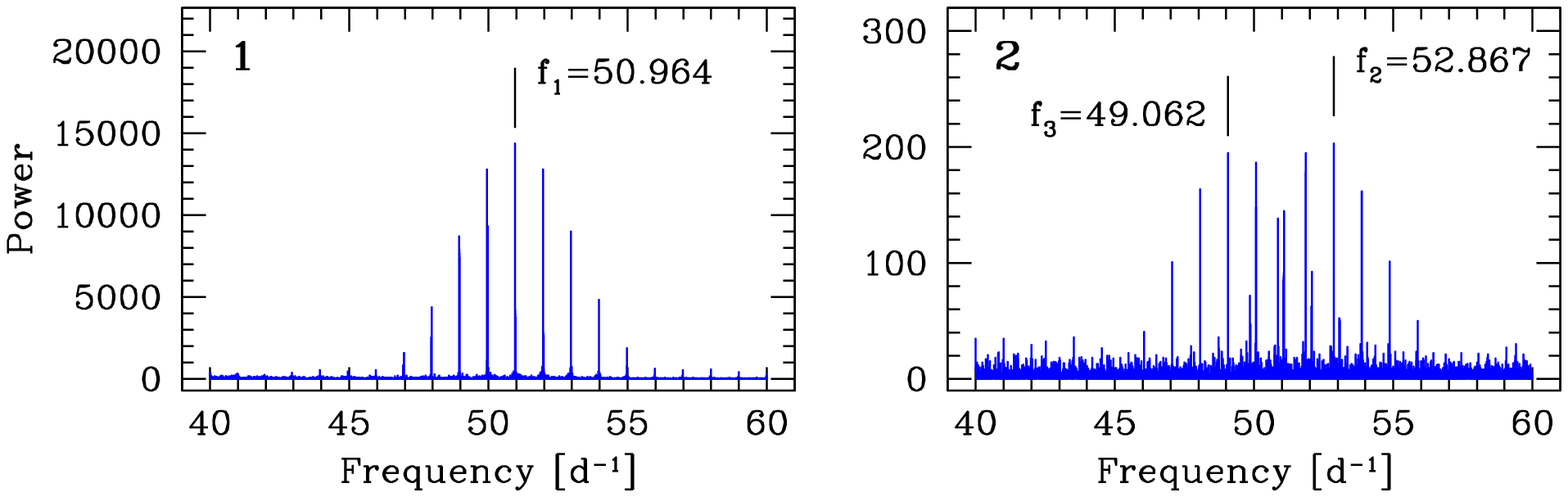}}
\FigCap{Power spectra of the short-period pulsator OGLE-GD-DSCT-0058
with labelled frequencies.}
\end{figure}

%%%%%%%%%%%%%%%%%%%%%%%%%%%%%%%%%%%%%%%%%%%%%%%%%%%%%%%%%%%%%%%%%%%%%

\Subsection{Other short-period pulsators}

In the observed Galactic fields, we detected 60 stars with periods
$<0.23$~d, $I$-band variability range $>0.05$~mag, and shapes characteristic
for pulsating stars. Fifty-two of the stars show more than one
periodicity, but the period ratios cannot be explained by the presence
of pure radial modes in $\delta$~Sct. In some of the stars we observe
close, very likely non-radial modes (see Fig.~13).
In Fig.~14, we present the $I$ \vs $V-I$ diagram for one of the OGLE
fields with marked locations of the short-period stars.
Most of the objects are probably $\delta$~Sct type stars, but we cannot
exclude that some of them are more distant $\beta$~Cep type stars.
Information on spectral type would help in the final classification
of these objects.

\begin{figure}[htb!]
\centerline{\includegraphics[angle=0,width=130mm]{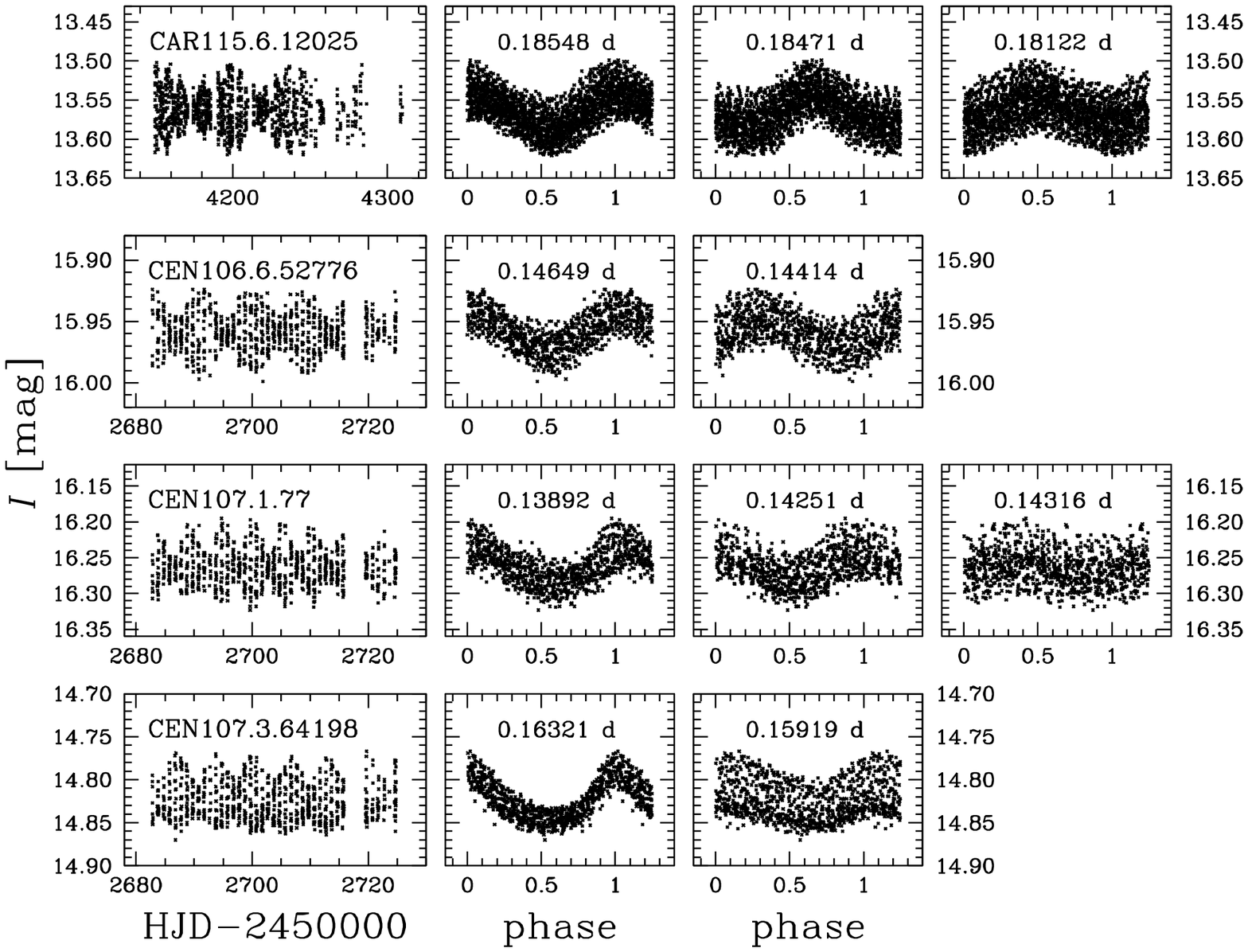}}
\FigCap{Example light curves of short-period pulsators showing
two or more close modes. These stars could be either $\delta$~Sct
or $\beta$~Cep type variables, but due to the lack of information
on intrinsic color or surface temperature it is almost impossible
to distinguish between the two types.}
\end{figure}

\begin{figure}[htb!]
\centerline{\includegraphics[angle=0,width=100mm]{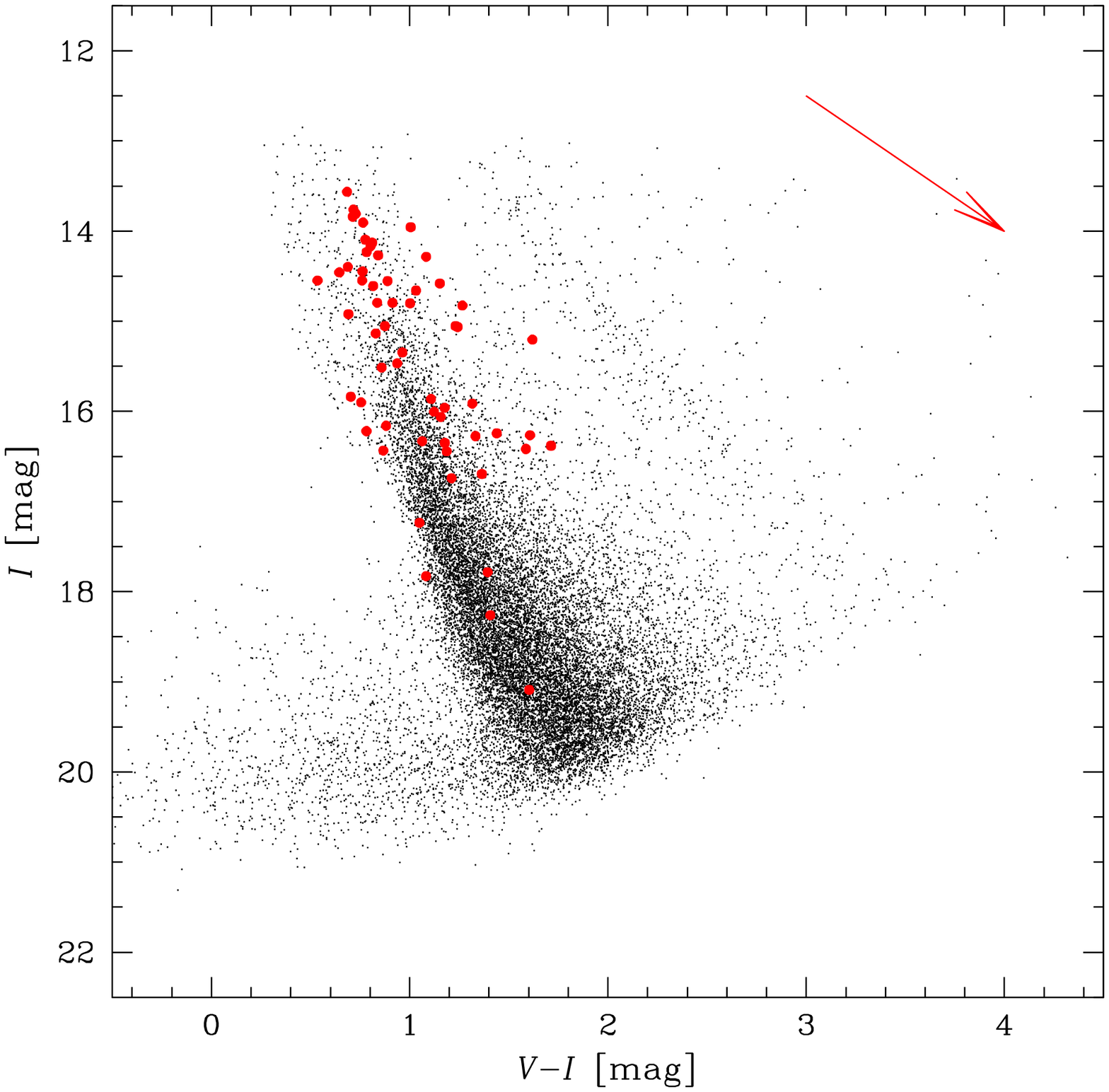}}
\FigCap{Color-magnitude diagram of the field CAR113.4 located about
$0\zdot\arcd9$ from the Galactic plane with the location of 58 unclassified
short-period pulsators (red points). Red arrow indicate the direction
of reddening with an assumed ratio of total-to-selective exinction
$R_I=A_I/E(V-I)=1.5$.}
\end{figure}

%%%%%%%%%%%%%%%%%%%%%%%%%%%%%%%%%%%%%%%%%%%%%%%%%%%%%%%%%%%%%%%%%%%%%

\Subsection{Pulsating white dwarf}

The high-frequency search (24--100~d$^{-1}$) led us to the initial detection
of nearly one hundred objects with a signal-to-noise ratio S/N$>10$.
All but one objects have $V-I>0.3$~mag. The blue outlier
with $V-I\approx-0.23$~mag has the period of 0.01273972(1)~d$=1100.712(1)$~s
and the full $I$-band amplitude of 0.010~mag. The observed color, period,
and amplitude are typical for pulsating white dwarfs, for instance of ZZ~Cet
type (\eg Eggen 1985, Van Grootel \etal 2012). Phased light curve
of OGLE-GD-WD-0001 is shown in Fig.~15.

\begin{figure}[htb!]
\centerline{\includegraphics[angle=0,width=60mm]{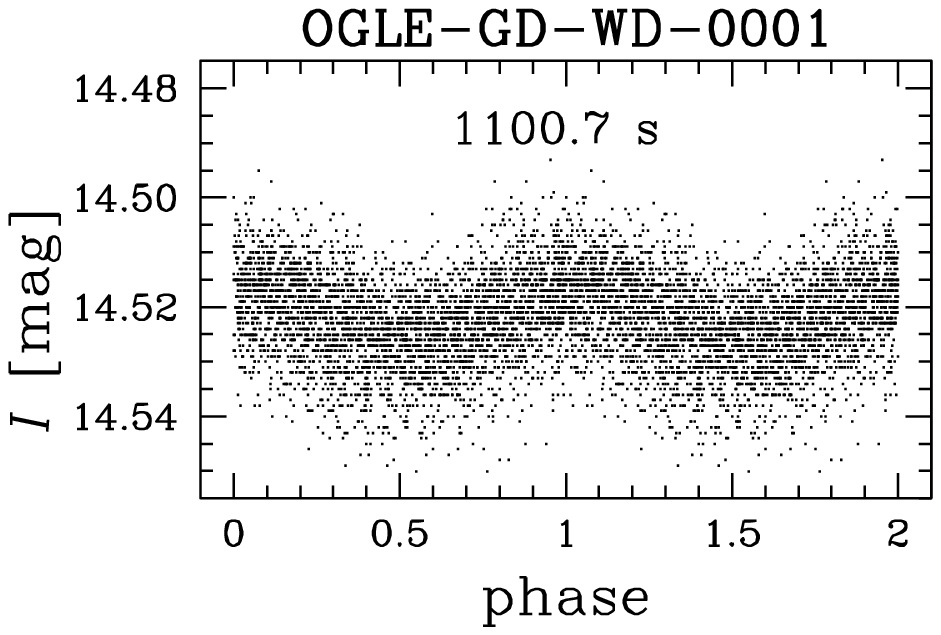}}
\FigCap{Light curve of the pulsating white dwarf detected in the
OGLE Galactic disk area.}
\end{figure}

%%%%%%%%%%%%%%%%%%%%%%%%%%%%%%%%%%%%%%%%%%%%%%%%%%%%%%%%%%%%%%%%%%%%

\Section{Completeness of the search}

The completeness of the search for pulsating stars depends on their type.
All our Classical Cepheids were detected during the inspection of stars
with a signal-to-noise ratio S/N$\geq14$. We did not find any new Cepheid
among objects with $10\leq$S/N$<14$. The discovered Cepheids are bright
and have the average $I$-band magnitudes $<16.5$. Given the brightness
and the corresponding {\it rms} variability $\sigma_I<0.03$~mag (Szyma\'nski
\etal 2010) it is highly unlikely we missed a single Cepheid. Therefore,
we conclude that the completeness of the survey for Classical Cepheids with
amplitudes $>0.1$~mag reached 100\%. A close to 100\% completeness also holds
for the RRab type stars of which only one has a S/N$<14$. Sinusoidal-like
RRc type objects are difficult to distinguish from hundreds of regularly
variable candidate spotted stars and some pulsators of this type could be
missed. Their completeness could be as low as 50\%.

For LPVs the completeness seems to depend on the time coverage of the fields.
We did not detect these variables in the fields with only one well-covered
season, \ie CAR106, CEN106, and CEN107. The new LPVs were found in the
disk fields which contain about 83\% of all monitored stars.
This percentage is a rough estimate of the completeness for high-amplitude
regularly pulsating LPVs, in particular the Mira variables.

Proper classification of short-period pulsators based on the $VI$
photometry is particularly difficult for stars in the Galactic disk fields.
However, the results for the fundamental-mode HADS variables seem
to be very complete, since 24 objects out of the 28 detected
fundamental-mode $\delta$~Sct stars were found during the inspection
of stars with the variability signal-to-noise ratio S/N$\geq14$.

%%%%%%%%%%%%%%%%%%%%%%%%%%%%%%%%%%%%%%%%%%%%%%%%%%%%%%%%%%%%%%%%%%%%%

\Section{Summary of the search for pulsating stars}

Twenty-one fields in the direction tangent to the Centaurus Arm of the
Galactic disk were observed by the OGLE survey during the third phase of the
project in years 2001--2009. The analysis of the data has brought an
identification of 20 {\it bona fide} Classical Cepheids, 45 RR~Lyr type
stars, 31 LPVs, one pulsating white dwarf, and 58 very likely $\delta$~Sct
type stars. Only one of the disk variables, OGLE-GD-CEP-0019, had been
identified before. Ten Cepheids are fundamental-mode, two are first-overtone,
and eight are double-mode pulsators, mainly F/1O and 1O/2O.
One Cepheid, OGLE-GD-CEP-0001, pulsates in the first overtone
and a second mysterious mode. All detected Classical Cepheids are
brighter than $I=16.4$~mag, which is about 5~mag above the survey limit.
We also report six candidates for Type II Cepheids.

Among the RR~Lyr variables, 36 stars are of the RRab type and nine of the
RRc type. The photometrically derived metallicities for RRab variables
indicate the presence of stars from two populations: metal poor stars of
the Galactic halo and metal rich stars of the thick disk.

Twenty-eight of the identified $\delta$~Sct type stars pulsate in the
fundamental mode. In 22 stars of this type we found two radial modes,
in six stars three radial modes, and in one object four radial modes.
Many stars exhibit additional non-radial modes. The presence of more than
two radial modes in seven stars allowed the determination of their distance
and physical parameters based on pulsation models. We show that two
triple-mode and one double-mode stars are very likely Population II
pulsators of SX~Phe type. We note that star OGLE-GD-DSCT-0012
cannot be reproduced by our models. It could be either a $\delta$~Sct
type star pulsating in some non-radial modes or, less probable,
a $\beta$~Cep type star. Information on spectral type would help
in the definitive classification of this and another 60 short-period
pulsating variables found in the OGLE disk area. Among the detected
stars there is also a candidate $\delta$~Sct star with an exceptionally
high full $I$-band amplitude of 0.35~mag at the very short period of 0.0196~d.

The detection completeness of Classical Cepheids, RRab type stars, and
fundametal-mode $\delta$~Sct type stars with amplitudes $>0.1$~mag
and brightness $I<18$~mag is assessed at the level of 100\%.
In the case of the RRc variables several stars could be missed.
The search for LPVs was estimated to be complete in about 83\%.

The OGLE photometry in the disk fields allowed a confident classification
of Classical Cepheids, RR~Lyr, most of LPVs and some $\delta$~Sct type variables.
However, many other types of pulsating variables, besides $\beta$~Cep type
stars, are very likely present in the data. Among them could be low-amplitude
Cepheids, SPB stars, $\gamma$~Dor type stars, hybrid $\beta$~Cep/SPB
(Pamyatnykh 1999), $\delta$~Sct/$\gamma$~Dor stars (Grigahc\`ene \etal 2010),
and pulsating M dwarfs (Rodr\'iguez-L\'opez, MacDonald and Moya 2012).
Similar periods, small amplitudes, and sinusoidal-like light curves
of the variables make them extremely difficult to discriminate from
candidate spotted stars.

\Section{The OGLE-IV Galactic Disk Survey}

The detection of new Classical Cepheids is one of the aims of a new Galactic
survey conducted in the framework of the OGLE-IV project at the 1.3-m
Warsaw telescope. The survey covers more than half of the Galactic plane
with longitudes from $-170\arcd$ to $+60\arcd$ and latitudes
$-3\arcd \lesssim b \lesssim +3\arcd$. By taking 30~s exposures
we plan to find variables in a brightness range $10 < I < 18$~mag.
In Fig.~16, we present a color image of almost entire fourth quarter
of the Galactic disk with the overlaid location of the OGLE-III
and OGLE-IV fields. The total OGLE-IV disk area is about 200 times
larger that the area analyzed in this paper. With this coverage
we plan to achieve a nearly complete census of high-amplitude Classical
Cepheids.

\begin{figure}[htb!]
\centerline{\includegraphics[angle=90,width=85mm]{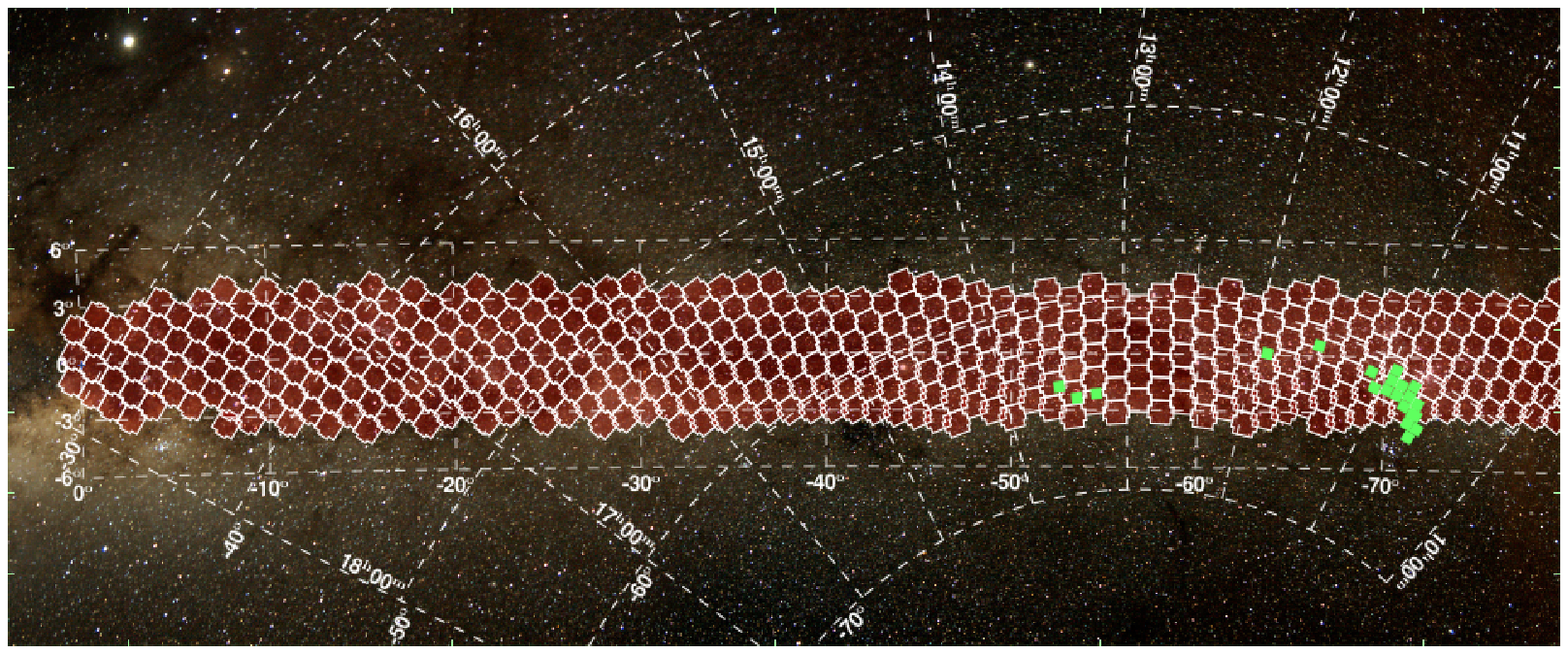}}
\FigCap{Map of the OGLE-IV Galactic Disk Survey fields with longitudes
between $-80\arcd$ and $0\arcd$. Dashed white lines mark the equatorial
and Galactic coordinate grids. The whole OGLE-IV disk survey covers
the Galactic plane between longitudes $-170\arcd\lesssim l\lesssim+60\arcd$
and latitudes $-3\arcd\lesssim b\lesssim+3\arcd$. Green filled squares
denote the location of the OGLE-III fields analyzed in this paper.}
\end{figure}

For the purpose of the new survey we compiled data on known Galactic
Classical Cepheids from the literature. The list of confirmed Classical Cepheids
from the General Catalogue of Variable Stars (GCVS, Samus \etal 2013)
was supplemented with the OGLE-III objects (this work and Soszy\'nski \etal 2011b),
revised objects from the All Sky Automated Survey (ASAS) Catalogue of Variable
Stars (Pojma\'nski 1998, 2002, 2003, Pojma\'nski and Maciejewski 2004, 2005,
Pojma\'nski, Pilecki and Szczygie{\l} 2005) and the International
Variable Star Index\footnote{http://www.aavso.org/vsx/}.
We used photometric data from the Northern Sky Variability Survey
(NSVS, Wo\'zniak \etal 2004) and the ASAS survey to estimate
the pulsation periods for some Cepheids. Based on the ASAS data
(G. Pojma\'nski, private communication) we found that object V2864~Sgr with
a period of 75.8~d is a LPV star, not a Cepheid. Currently, the longest period
Classical Cepheid known in our Galaxy is S~Vul with $P=68.464$~d.
The compiled list of 841 Galactic Classical Cepheids is also available
in the on-line OGLE archive (see links in Section~3).

Fig.~17 shows the distribution of the Galactic Classical Cepheids
in the ($l$,$b$) coordinates. Approximately 85\% of the stars
are located within a $-5\arcd < b < +5\arcd$ stripe and $\approx64$\%
within a $-3\arcd < b < +3\arcd$ stripe along the Galactic plane.
From the distribution we can infer that obscured Milky Way plane regions
in the first and fourth quarters ($-90\arcd \leq l < +90\arcd$)
may hide many (probably at least several hundreds) not yet detected Cepheids.

\begin{figure}[htb!]
\centerline{\includegraphics[angle=0,width=130mm]{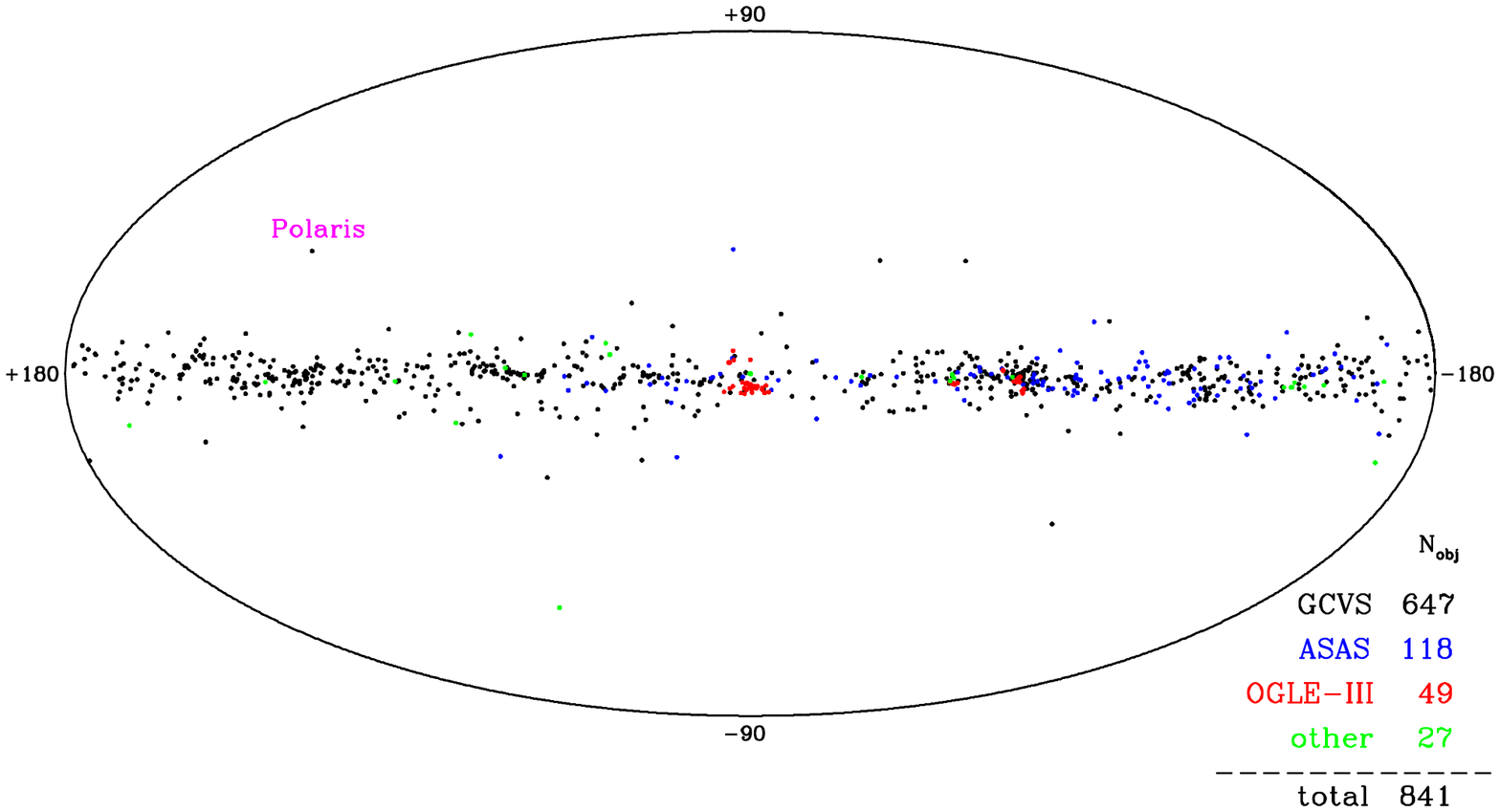}}
\FigCap{Distribution of the Milky Way's Classical Cepheids in Galactic
coordinates in equal area Mollweide{\rm '}s projection.}
\end{figure}

In Fig.18, we present the histogram of the pulsation periods of known
funda-mental-mode Galactic Classical Cepheids and compare it with period
distributions for stars in the SMC (Soszy\'nski \etal 2010) and LMC
(Soszy\'nski \etal 2008, Ulaczyk \etal 2013). The distribution
for the Galactic Cepheids peaks at $\approx5.0$~d, while for LMC and SMC
at $\approx3.0$~d and $\approx1.5$~d, respectively. This shift is
a result of different metallicities and chemical evolution
between the Galaxy, LMC, and SMC. A future more complete list of Galactic
Classical Cepheids from OGLE-IV will be used to trace the spiral
structure and star formation history of the Milky Way.

\begin{figure}[htb!]
\centerline{\includegraphics[angle=0,width=130mm]{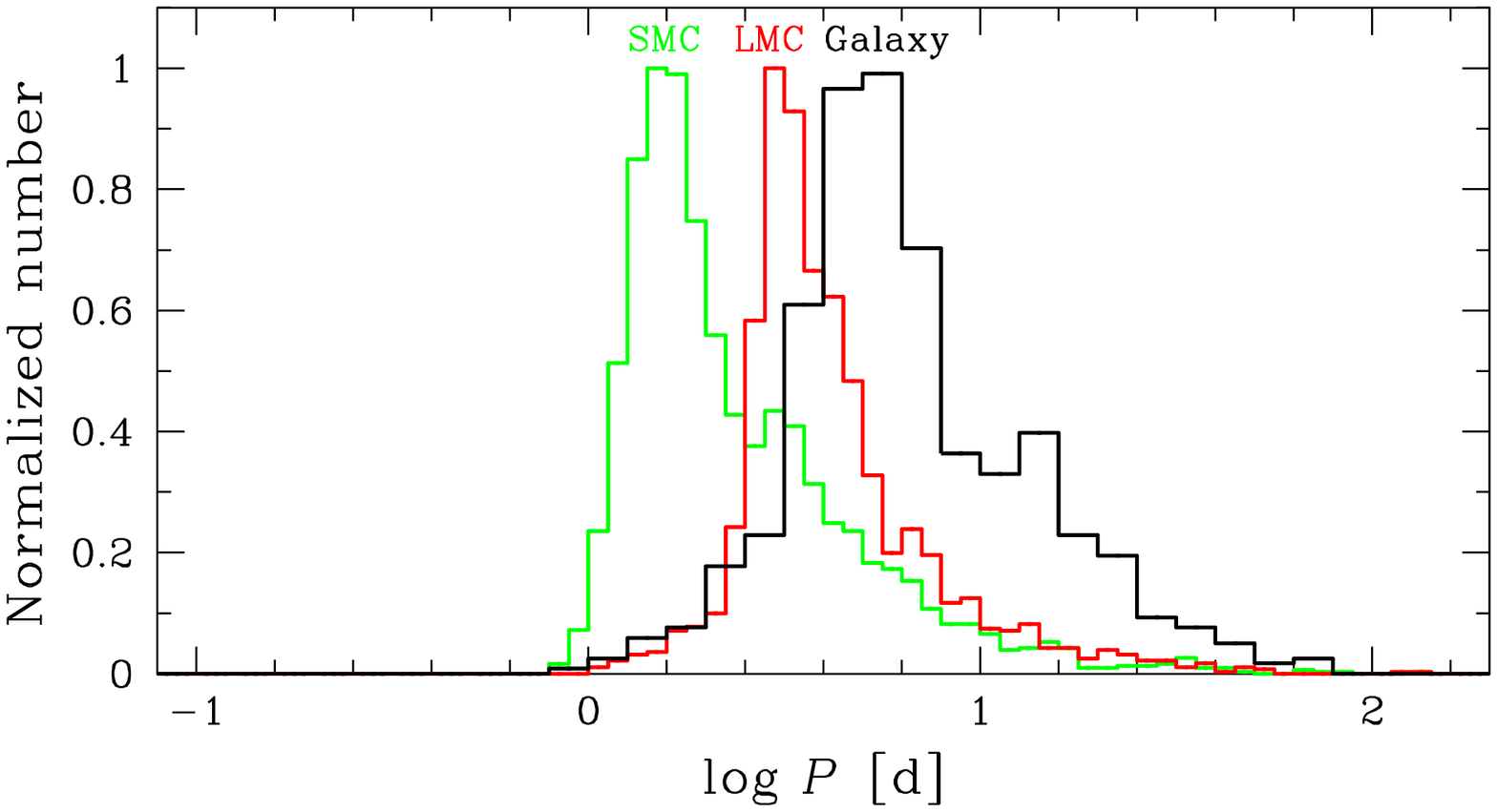}}
\FigCap{Comparison of the period distributions for fundamental-mode
Classical Cepheids in the SMC, LMC, and our Galaxy
(2626, 1851, and 664 objects, respectively).}
\end{figure}

%%%%%%%%%%%%%%%%%%%%%%%%%%%%%%%%%%%%%%%%%%%%%%%%%%%%%%%%%%%%%%%%%%%%%

\Acknow{
The authors would like to thank Richard I. Anderson for discussions
on the presented project, Grzegorz Pojma\'nski for his help in close
inspection of selected ASAS objects, and Thomas Fruth for providing
the BEST-II survey photometric data on candidates for Galactic Cepheids.
This work has been supported by the Polish Ministry of Sciences
and Higher Education grants No. IP2012 005672 under the Iuventus Plus
program and No. IdP2012 000162 under the Ideas Plus program.
W.D. is supported by the Polish National Science Centre through
grant No. DEC-2012/05/B/ST9/03932.
The OGLE project has received funding from the European
Research Council under the European Community$'$s Seventh Framework
Programme (FP7/2007-2013)/ERC grant agreement No. 246678 to A.U.}

%%%%%%%%%%%%%%%%%%%%%%%%%%%%%%%%%%%%%%%%%%%%%%%%%%%%%%%%%%%%%%%%%%%%%

\end{document}